\documentclass[twocolumn, tightenlines, amsmath, amssymb, superscriptaddress]{revtex4-1}

\usepackage{graphicx}
\usepackage{amsmath}
\usepackage{amssymb}
\usepackage{epstopdf}
\usepackage{color}
\usepackage{lipsum}
\usepackage{ulem}
\usepackage[dvipsnames]{xcolor}
\usepackage[colorlinks=true]{hyperref} 
\usepackage{braket}
\usepackage{lineno}
\usepackage{tikz}
\usepackage{subcaption}
\usepackage{siunitx}

\usetikzlibrary{quantikz}
\usepackage[separate-uncertainty=true]{siunitx}
\usepackage[colorlinks=true]{hyperref}
\usepackage{mathtools, cuted}
\usepackage{braket}
\newcommand{\ketbra}[2]{\mathinner{|{#1}\rangle\langle{#2}|}}

\makeatletter
\newcommand\footnoteref[1]{\protected@xdef\@thefnmark{\ref{#1}}\@footnotemark}
\makeatother

\definecolor{qdarkgray}{HTML}{171710} 
\definecolor{qblack}{HTML}{000000}
\definecolor{qwhite}{HTML}{ffffff}
\definecolor{qlightgray}{HTML}{6e7075}
\definecolor{qgray}{HTML}{383e48}
\definecolor{qpurple1}{HTML}{946cba}
\definecolor{qpurple2}{HTML}{8888c6}
\definecolor{qblue1}{HTML}{6ab5db}
\definecolor{qblue2}{HTML}{7aa0d2}
\definecolor{qblue3}{HTML}{58c4e1}

\hypersetup{colorlinks=true,
            linkcolor=qblue2,
            citecolor=qpurple1,
            urlcolor=qlightgray}

\begin{document}

\author{Nicolas Maring}
\author{Andreas Fyrillas}
\thanks{These authors contributed equally to this work.}
\affiliation{Quandela, 7 Rue Léonard de Vinci, 91300 Massy, France}
\author{Mathias Pont}
\thanks{These authors contributed equally to this work.}
\affiliation{Quandela, 7 Rue Léonard de Vinci, 91300 Massy, France}
\affiliation{Centre for Nanosciences and Nanotechnologies, CNRS, Universit\'e Paris-Saclay, UMR 9001, 10 Boulevard Thomas Gobert, 91120, Palaiseau, France}
\author{Edouard Ivanov}
\thanks{These authors contributed equally to this work.}
\affiliation{Quandela, 7 Rue Léonard de Vinci, 91300 Massy, France}
\author{Petr Stepanov}
\author{Nico Margaria}
\author{William Hease}
\author{Anton Pishchagin}
\author{Thi Huong Au}
\author{Sébastien Boissier}
\author{Eric Bertasi}
\author{Aurélien Baert}
\author{Mario Valdivia}
\author{Marie Billard}
\author{Ozan Acar}
\author{Alexandre Brieussel}
\author{Rawad Mezher}
\author{Stephen C.\ Wein}
\author{Alexia Salavrakos}
\author{Patrick Sinnott}
\affiliation{Quandela, 7 Rue Léonard de Vinci, 91300 Massy, France}
\author{Dario A.\ Fioretto}
\affiliation{Centre for Nanosciences and Nanotechnologies, CNRS, Universit\'e Paris-Saclay, UMR 9001, 10 Boulevard Thomas Gobert, 91120, Palaiseau, France}
\author{Pierre-Emmanuel Emeriau}
\affiliation{Quandela, 7 Rue Léonard de Vinci, 91300 Massy, France}
\author{Nadia Belabas}
\affiliation{Centre for Nanosciences and Nanotechnologies, CNRS, Universit\'e Paris-Saclay, UMR 9001, 10 Boulevard Thomas Gobert, 91120, Palaiseau, France}
\author{Shane Mansfield}
\affiliation{Quandela, 7 Rue Léonard de Vinci, 91300 Massy, France}
\author{Pascale Senellart}
\affiliation{Centre for Nanosciences and Nanotechnologies, CNRS, Universit\'e Paris-Saclay, UMR 9001, 10 Boulevard Thomas Gobert, 91120, Palaiseau, France}
\author{Jean Senellart}
\author{Niccolo Somaschi}
\affiliation{Quandela, 7 Rue Léonard de Vinci, 91300 Massy, France}

\title{A general-purpose single-photon-based quantum computing platform}

\date{\today}

\begin{abstract}
Quantum computing aims at exploiting quantum phenomena to efficiently perform   computations that are unfeasible even for the most powerful classical supercomputers. Among the promising technological approaches, photonic quantum computing offers the advantages of low decoherence, information processing with modest cryogenic requirements, and native integration with classical and quantum networks. To date, quantum computing demonstrations with light have implemented specific tasks with specialized hardware, notably Gaussian Boson Sampling which permitted quantum computational advantage to be reached.  Here we report a first  user-ready general-purpose quantum computing prototype based on single photons. The device comprises a high-efficiency quantum-dot single-photon source feeding a universal linear optical network on a reconfigurable chip for which hardware errors are compensated by a machine-learned transpilation process. Our full software stack allows remote control of the device to perform computations via logic gates or direct photonic operations.
For gate-based computation we benchmark one-, two- and three-qubit gates with state-of-the art fidelities of $99.6\pm0.1 \%$, $93.8\pm0.6 \%$ and $86\pm1.2 \%$ respectively. We also implement a variational quantum eigensolver, which we use to calculate the energy levels of the hydrogen molecule with high accuracy. For photon native computation, we implement a classifier algorithm using a $3$-photon-based quantum neural network  and report a first $6$-photon Boson Sampling demonstration on a universal reconfigurable integrated circuit. Finally, we report on a first heralded 3-photon entanglement generation, a key milestone toward measurement-based quantum computing.

\end{abstract}

\maketitle

 Realizations of quantum computing (QC) have built on rapid progress in controlling physical systems that can support quantum information such as superconducting circuits (e.g.~\cite{arute2019, zhu2022}), trapped ions (e.g.~\cite{moses2023race,debnath2016demonstration}), neutral atoms (e.g.~\cite{bluvstein2022quantum}) and light (e.g.~\cite{zhong2021, arrazola2021}).
These technological breakthroughs have brought four platforms to the regime of quantum computational advantage~\cite{arute2019,zhu2022,wu2021strong,madsen2022,zhong2020}, by solving specific sampling problems that would require unreasonable computing time even for the most powerful classical supercomputers.
Two of these four are photonic, highlighting the position of light-based technology among the leading platforms.
Quantum light as quantum information carrier offers the advantage of low decoherence and comes with a large choice of degrees of freedom to encode the information, while providing natural connectivity for distributed or blind quantum computing \cite{broadbent2009universal}.

\begin{figure*}[]
    \centering
    \includegraphics[width=0.90\linewidth]{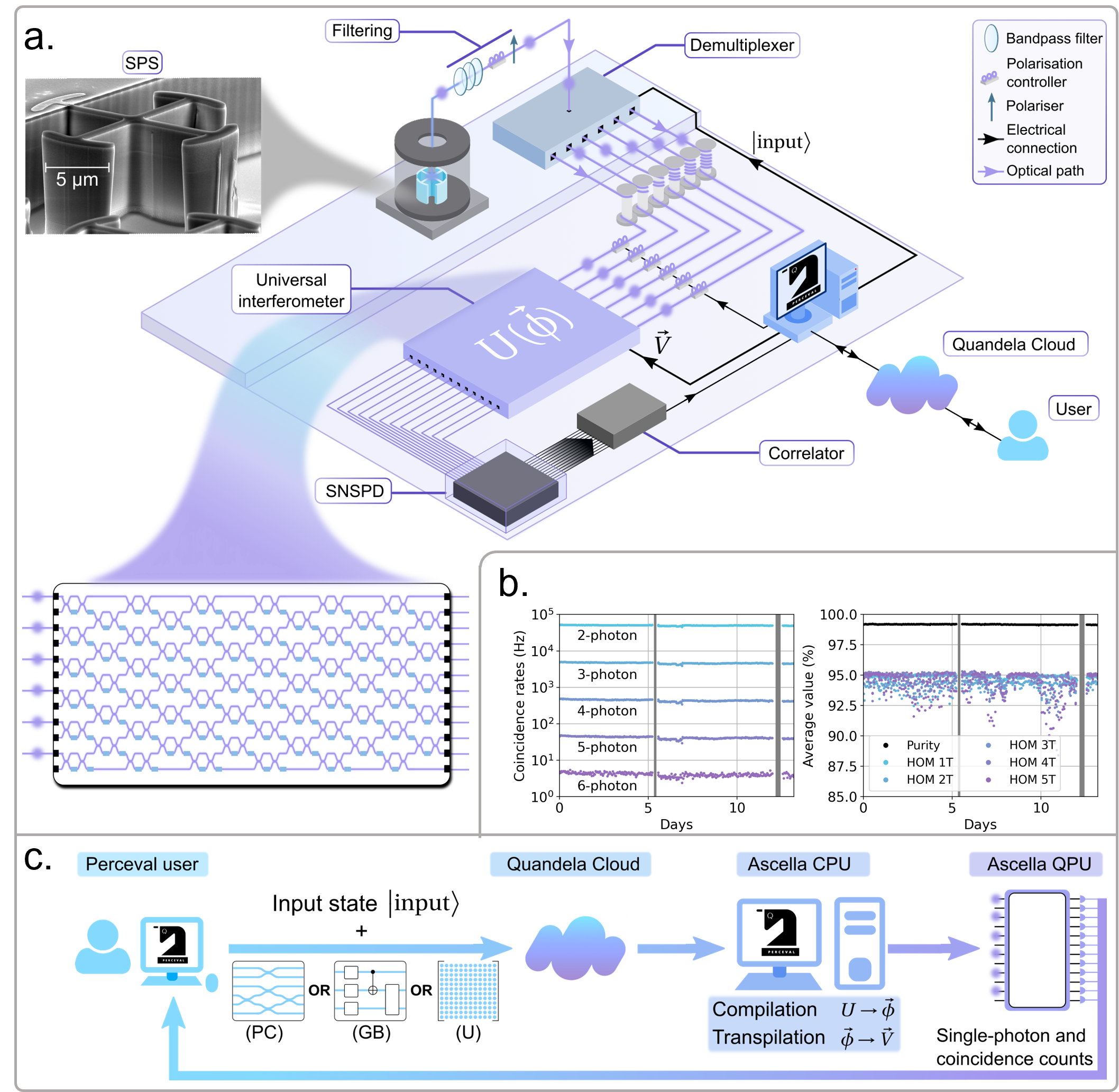}
    \caption{\textbf{Architecture, performance and stability of Ascella.} 
    \textbf{a.} Sketch of the overall architecture of the 6 single-photon quantum computer. A  quantum-dot single-photon source (SPS) device at $5$K is operated at $80$ MHz repetition rate. An active demultiplexer followed by fibered delays converts the train of single photons into $6$ photons arriving simultaneously to the universal 12-mode photonic chip. Photons are detected at the chip output by superconducting nanowire single-photon detectors (SNSPD) and detection times are processed by a correlator.
    A full software stack controls the unitary matrix $U$ implemented on the chip through the voltages $\vec{V}$ applied on $126$ thermal phase shifters, yielding phase shifts $\vec{\phi}$,  and the photonic input state according to the job requested. It also recalibrates hourly and readjusts all hardware control knobs for optimal performance. The single photons are sent into a photonic chip featuring a universal interferometer scheme capable  of implementing any $12\times12$ unitary matrix. \textbf{b.} Detected $N$-photon coincidence rates for $N$-photon inputs as a function of time, with the photonic circuit configured to implement the identity matrix. Rates are integrated for $50$ seconds. Grey areas correspond to maintenance and upgrade periods. In right figure, we also monitor the on-chip photon indistinguishability and single-photon purity  quantified respectively by the Hong-Ou-Mandel (HOM) visibility $V_\text{HOM}$ and $1-g^{(2)}(0)$, where $g^{(2)}$ is the normalized second-order correlation function. HOM $k$T is $V_\text{HOM}$ for delays $k\times\Delta T$  between emitted photons where $\Delta T=\qty{180}{\nano\second}$. Each data point corresponds to a  correlation histogram integrated over 10 seconds.  
    \textbf{c.} Job execution flowchart on Ascella. \textit{Perceval} users may send jobs consisting in photonic circuits (PC), or a gate-based circuit (GB), or a unitary matrix (U), along with the desired input state to the Quandela Cloud. The job is first processed by a CPU, which computes the necessary phase shifts $\vec{\phi}$ to apply, and subsequently the voltages $\vec{V}$ for the on-chip phase shifters from our compilation and transpilation process. Finally, the user receives the collected single-photon and coincidence counts after the computation on the quantum processing unit (QPU).
    }
    \label{fig:main_hardware}
\end{figure*}

Over the years, a variety of proposals for  universal fault-tolerant computing  have been put forward in the discrete-variable photonic approach in which quantum information is encoded with single photons~\cite{raussendorf2001, li2015, herr2018, auger2018, bartolucci2023}.
With identified thresholds, these roadmaps motivate the development of quantum computing hardware based on single-photon sources, integrated photonic chips and single-photon detectors.
Experimental progress of ever increasing complexity has been achieved with integrated sources exploiting nonlinear effects, including with large-scale integrated chips~\cite{vigliar2021, bao2023}.
However, the probabilistic nature of the single-photon generation process, the need for it to be heralded and the requirement to operate at low efficiency to limit multiphoton events are strong constraints on the hardware architecture.
This has resulted in a limited  number of manipulated photons  with typical rates in the mHz range for $4$ photons and the demonstration of specific information processing tasks relying on dedicated photonic chips~\cite{bao2023}. Overcoming these limitations is foreseen through the use of massive multiplexing of hundreds of heralded sources~\cite{bombin2021interleaving}.

Another path to large-scale QC with single photons has progressively emerged owing to deterministic single photon source devices based on semiconductor quantum dots (QDs)~\cite{somaschi2016, ding2016, wang2019, uppu2020, tomm2021}. Such sources have demonstrated record single-photon generation efficiency, 10-20 times higher than their nonlinear conterparts, allowing for a drastic reduction in resource requirements. Such efficiencies allowed a record manipulation of $14$ single photons in a free-space Boson Sampling experiment \cite{wang2019boson}. Very recently, the same QD sources have shown their ability to deterministically generate photonic cluster states at high rate \cite{coste2022high}, reducing even further the foreseen overheads for large-scale measurement-based quantum computation~\cite{gimeno2015}.

In the present work, we report on the first general-purpose user-ready single-photon-based quantum computing machine, named Ascella.  It is cloud-accessible~\cite{QuandelaCloud} and based on six photonic qubits generated by an on-demand QD source. 
The  quantum information is encoded in the path degree of freedom  and arbitrarily manipulated in a 12-mode integrated universal interferometer. A machine-learned transpilation process  corrects for the hardware manufacturing errors. Ascella operates the largest number of single photons on chip to date with a 6-photon sampling rate of 4 Hz  and shows operation stability over weeks. We benchmark its performances and demonstrate applications both in the gate-based and photonic computation frameworks. Each reported result represents either state-of-the-art performance or first-ever experimental demonstration for which we provide the full code to reproduce through Quandela Cloud.
The numerous applications illustrate the general-purpose potential of the machine for noisy near-term quantum computing.
% The multitude of applications demonstrates the general purpose of the machine for noisy intermediate scale quantum computing.
We finally discuss the evolution of the reported platform towards scale-up, and demonstrate for the first time a critical step for future measurement-based quantum computation: heralded entanglement generation of three-photon GHZ states. 

\section*{Single-photon based computer }

\subsection{Architecture}
Ascella's hardware, as shown in Fig.\,\ref{fig:main_hardware}.a, comprises an on-demand high-brightness single-photon source, a programmable optical demultiplexer allowing up to 6 single photons to simultaneously interfere on a 12-mode reconfigurable universal interferometer, and a single-photon detection and post-processing unit. 

The on-demand single-photon source (see Supplementary~\ref{app:SPS}) based on an InGaAs quantum dot in a microcavity~\cite{somaschi2016} is optically excited at an \qty{80}{\mega\hertz} rate. It exploits a neutral dot and LA-phonon-assisted near-resonant excitation~\cite{thomas2021} to emit linearly-polarized single photons with $55 \%$ probability into the collection lens.  To send $6$ single photons to every even input mode of the chip, an active optical demultiplexer sequentially deflects the photon stream into $6$ optical fibers of different lengths adjusted to synchronize the photons~\cite{pont2022}. Using optical shutters, the demultiplexer can prepare any input state from $\ket{000000000000}$ to $\ket{101010101010}$ (photon positions can subsequently be swapped, see Supplementary~\ref{app:setup}). The $12$-mode photonic integrated circuit (Si$_3$N$_4$ platform) is composed of $126$ voltage-controlled thermo-optic phase shifters and $132$ directional couplers~\cite{taballione2020} which are laid out in a rectangular universal interferometer scheme (see Fig.\ref{fig:main_hardware}.a). 
Finally, the $12$ outputs of the circuit are connected to high-efficiency superconducting nanowire single-photon detectors (SNSPD), and $N$-photon detection events are registered using a time-to-digital converter. 

The average total efficiency of the optical setup is $\sim8\,\%$, including the single-photon source device brightness, transmission of all optical components, and detection efficiencies (see Supplementary~\ref{app:setup}). This leads to record-breaking $2$- to $4$-photon on-chip coincidence rates (Fig.\,\ref{fig:main_hardware}.b) and the first on-chip processing of $5$ and $6$ photons, at respective rates of $50$~Hz  and $4$~Hz. 
We measure high single-photon purity $>99\,\%$, high indistinguishability $\sim 94\,\%$ independent of the delays between photon emission (up to $1\,\mu$s), resulting in a measured on-chip $2$-photon interference visibility for all $15$ pairs of $91-94\,\%$ (see Supplementary~\ref{app:indistinguishability}). The genuine $4$- and $6$-photon indistinguishability defined as the probability that the $N$ photons are identical establishes a new record value of $0.85\pm0.02$ for $4$ photons~\cite{pont2022quantifying}, and a first study for $6$ photons with value of $0.76\pm0.02$. We ensure long-term stability and high-performance operation of Ascella by monitoring key metrics and by carrying out automated hourly system optimization runs. This guarantees a highly-stable and long-term operation over several weeks (see Fig.\ref{fig:main_hardware}.b), robust against external temperature fluctuations and mechanical perturbations. 

To operate the machine, tasks are sent remotely with the python-based open-source framework %for programming photonic quantum computers
\textit{Perceval}~\cite{heurtel2023perceval}. The user can either specify (see Fig.\,\ref{fig:main_hardware}.c) a photonic circuit (PC), a gate-based circuit (GB) or a unitary transformation (U) to be applied to a specified input state containing $1$ to $6$ photons, and optional postselection criteria. Output photon coincidences are then acquired up to the desired sample number and data sample results are sent back to the user either as a stream of events or as an aggregated {\tt state:count} inventory.

\subsection{Chip control}
\label{sec:chip_control}

The rectangular universal interferometer layout (see Fig.\,\ref{fig:main_hardware}.a) is ideally based on balanced directional couplers (i.e.\ $\SI{50}{\%}$ reflectivity). Experimentally, we observe reflectivities with average values of $\SI{56.7(6)}{\%}$ for our chip at the operation wavelength of $928$~nm. The systematic error stemming from the fabrication process and wavelength dependency. These errors reduce the range of implementable $12\times12$ unitary matrices \cite{Burgwal2017, Russell2017} and, if not compensated for, affect the fidelity of the implemented unitary matrix to the target unitary matrix. To address these limitations, we designed a custom compilation and transpilation process that converts with high fidelity 
user-provided photonic circuits, unitary matrices or gate-based circuits into interferometer phase shift values (compilation) then into voltages to apply on the chip phase shifters (transpilation). Initially, a global optimization step fine-tunes the phase shifts to enhance matrix fidelity. Subsequently, the process calculates the voltages to apply on the chip phase shifters while compensating for thermal crosstalk. 
The thermo-optic phase shifters can be modelled by $\vec{\phi} = A\vec{V}^{\odot2}+\vec{b}$ where the vector $\vec{\phi}$ contains all $126$ physical phase shifts, $\vec{V}$  the $126$ applied voltages and $^{\odot2}$ represents element-wise squaring. Off-diagonal elements of the $126 \times 126$ matrix $A$ represent thermal crosstalk between phase shifters. We engineered a machine learning-based process that optimizes the values of $A$ and $\vec{b}$, constituting more than $16 000$ free parameters to determine. The same process also estimates individual directional coupler reflectivities and relative output losses (see Supplementary~\ref{app:chip_charac} for values).  This process offers a $7$-fold improvement on the transpilation ($\vec{\phi}$ to $\vec{V}$ process) over more standard characterization techniques involving interference-fringe measurements (see Methods). The full compilation and transpilation processes achieve an average fidelity of $F=99.7\pm0.08$ following the fidelity evaluation procedure from Ref.~\cite{taballione2020}.\\

\section*{Gate-based quantum computation }

\begin{figure*}[t]
    \centering
    \includegraphics[width=0.95\linewidth]{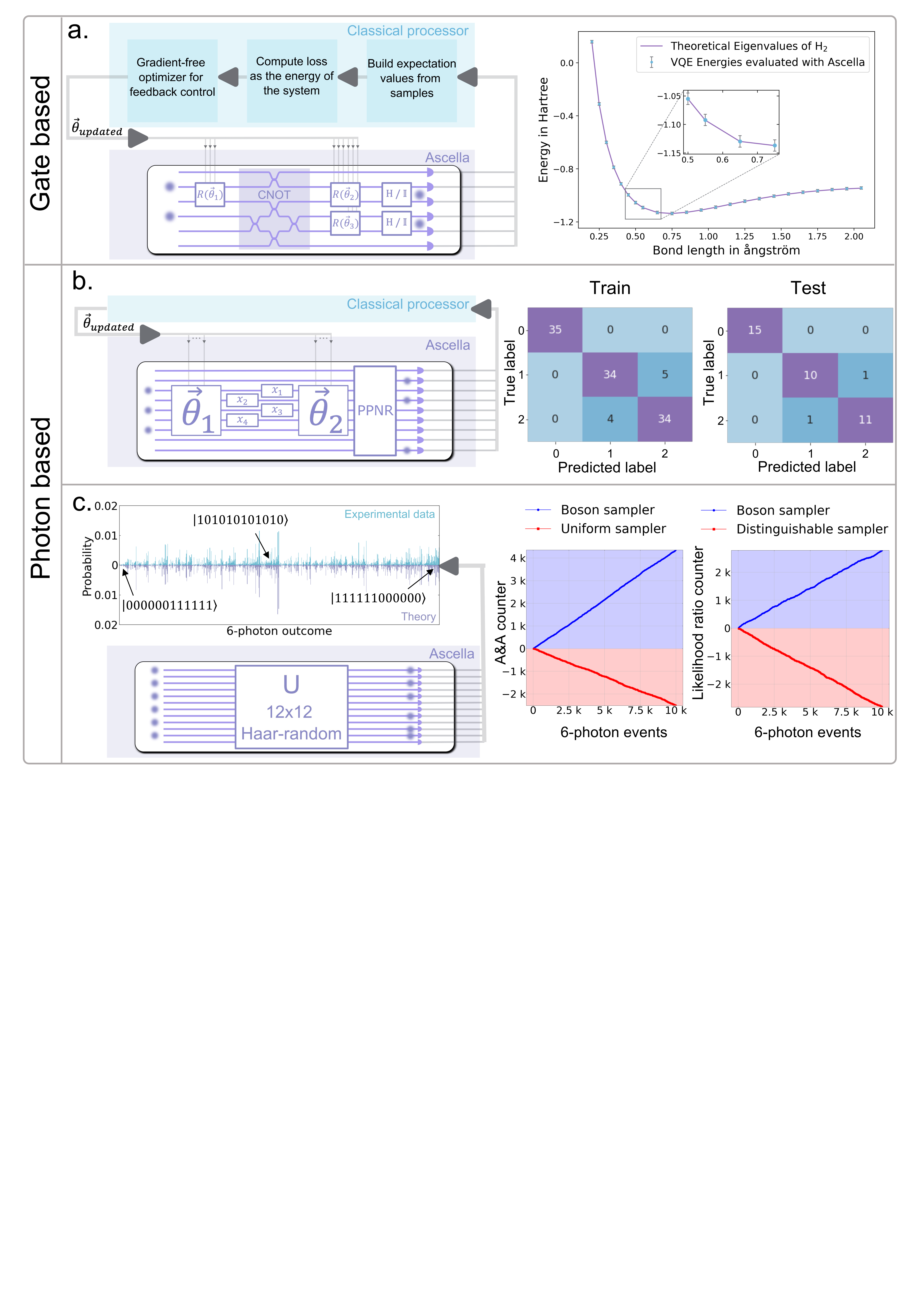}
    \caption{\textbf{a. Gate-based computation}. Hybrid variational quantum eigensolver. On Ascella, the single-qubit gates $R(\vec{\theta}_i)$, together with a CNOT gate, create an ansatz $2$-qubit state. We then measure in the Z basis (using the identity gate $\mathbb I$) or in the X basis (using the Hadamard gate H).
    The output counts (grey arrow) are sent to a classical processor which reconstructs the corresponding energy and implements a feedback loop to update the single-qubit gate angles $\vec \theta_i$ via a gradient-free optimizer and find an ansatz closer to the ground state. Each iteration on the QPU takes about $22$~s (including $14$~s of QPU time and classical communication to the cloud). The error bars are computed assuming a shot-noise limited error on the $2$-photon coincidences. \textbf{b.-c. Photon-native computation}. \textbf{b.} Classification task using a quantum neural network.  Confusion matrices for the classification of the IRIS dataset on Ascella: training dataset (left), test dataset (right). The accuracy is $0.92$ for the training set and $0.95$ for the test set. \textbf{c.} $6$ single-photon Boson Sampling. Measured (top) and modelled (bottom) $6$-photon output distributions for the input state $\ket{101010101010}$. The $924$ $6$-photon outcomes are canonically ordered from $\ket{000000111111}$ to $\ket{111111000000}$. Discrimination between Boson Sampling and uniform sampling hypothesis using the Aaronson and Arkhipov (A \& A) counter and between Boson Sampling and distinguishable sampling hypothesis using the likelihood ratio counter. The value of each discriminator is updated every $10^9$ samples, which corresponds to $\sim20$ $6$-photon events. In both cases, a positive slope validates the test (see Methods).} 
    \label{fig:applications}
\end{figure*}

Following the KLM scheme \cite{knill2001}, Ascella can perform probabilistic gate-based protocols. Within this quantum computation framework, we benchmark quantum logic gates on up to three qubits and implement a hybrid variational quantum  eigensolver.

\label{sec:benchmarking}

\subsection {Benchmarking logic gates} 
Ideally, a  gate $U$ applied to an initial pure state $\ket{\psi}$ will produce the pure state $U\ket{\psi}$. In reality,  errors, quantified by a noise channel $\Lambda$ \cite{nielsen2002quantum}, corrupt the final state, which is then described by a density matrix $\rho = \Lambda(U\ket{\psi}\!\bra{\psi}U^\dagger)$. A standard figure of merit to quantify the gate performance is the quantum state fidelity $F_{\psi}(U) = \braket{\psi|U^\dagger\rho U|\psi}$ of the final state $\rho$ to the  ideal state $U\ket{\psi}$. To assess Ascella's performance for a given gate, we evaluate the fidelity of the gate averaged over all possible input states $\ket{\psi}$, i.e.\ $F_\text{avg}(U) = \int F_{\psi}(U)d\psi$, where the integral is taken over the Haar measure \cite{collins2006integration}.

A brute-force approach to estimating $F_\text{avg}(U)$ requires an impractically large  number of measurements. A more efficient method, randomized benchmarking, has been proposed for matter qubits \cite{magesan2012characterizing} but applies long sequences of  gates from specific sets of unitaries 
\cite{dankert2009exact}. Since photonic quantum processing converts any quantum circuit to a photonic circuit \cite{clement2022lov}, we use a new method to evaluate $F_\text{avg}$~\cite{MW23}. Our method exploits symmetries so that the contribution of most $F_\psi$s to $F_\text{avg}$ cancel out, allowing $F_\text{avg}$ to be expressed as a finite discrete sum $F_\text{avg} = \sum_{i=1}^Kw_im_i$ of $K$  expectation values~$m_i$ with weight $w_i$ (see Supplementary~\ref{app:benchmarking}).
The $w_i$ and the state preparation and measurement configurations for each $m_i$  depend on the gate $U$ and are pre-computed. Each configuration consists in preparing an unentangled initial state $\ket{\psi}$, applying the gate and performing single-qubit Pauli measurements. For the gates benchmarked on Ascella (see Table \ref{table:gatebenchmarking}), the $K$ expectation values $m_i$ are obtained from $M \leq K$ measurement configurations, with $K$ less than the $\sim 2^{4n}$ measurements required for full process tomography \cite{mohseni2008quantum} of an $n$-qubit gate. 

The average gate fidelities measured for a $T$-gate defined as $T:=|0 \rangle \langle 0|+e^{i \frac{\pi}{4}}|1 \rangle \langle 1|$ \cite{bravyi2005universal}, a CNOT gate, and a Toffoli gate are shown in Table \ref{table:gatebenchmarking}. These measurements set a first benchmark for universal photonic quantum computing and are on par with the benchmarked performance of open-access quantum computing platforms based on ions and superconducting qubits (see Supplementary~\ref{app:benchmarking}). These  values are a lower bound on the true average gate fidelities since they also include errors related to  state preparation and measurement (SPAM) roughly given by $(1-F_\text{avg}(T\text{-gate})^{2n/3})$, which is $0.3\pm 0.1\%$ for the $T$-gate, $0.5\pm0.1\%$ for the CNOT gate, and $0.8
\pm 0.2\%$ for the Toffoli gate.

\begin{table}[]
    \centering
    \begin{tabular}{c|c|c|c|c|c}
        Qubits, $n$ & Gate, $U$ & $F_\text{avg}(U)$ ($\%$) & $M$ & $K$ & $2^{4n}$ \\
       \hline
        1 & $T$-gate & $99.6 \pm 0.1$ & 4 & 4 & 16\\
        2 & CNOT & $93.8\pm 0.6$ & 36 & 58 & 256\\
       3 &  Toffoli & $86\pm 1.2$ & 340 & 593 & 4096
    \end{tabular}
    \caption{ Average gate fidelities of $1$-, $2$- and $3$-qubit gates implemented by Ascella evaluated based on $K$ expectation values obtained from $M$ measurement configurations. }
    \label{table:gatebenchmarking}
\end{table}

\subsection{Variational quantum eigensolver}

We illustrate gate-based computation possibilities by implementing a variational quantum eigensolver (VQE) algorithm to compute the ground state energies of an H$_{2}$ molecule. VQE exploits the variational principle stating that given a Hamiltonian $\hat{\mathcal{H}}$ and an ansatz wavefunction $\ket {\psi (\vec \theta)}$ parameterized by $\vec \theta$, the ground state energy associated with $\hat{\mathcal{H}}$ satisfies $E_0 \leq \bra {\psi (\vec \theta)} \hat{\mathcal{H}} \ket {\psi (\vec \theta)}$ \cite{peruzzo2014variational,tilly2022variational}. In this context, VQE explores the state space by minimizing the energy to find a good approximation of $E_0$.

We build the fermionic Hamiltonian for H$_2$ using the symmetry-conserving Bravyi-Kitaev transformation \cite{bravyi2017tapering}, which is available through the \textit{OpenFermion} \cite{mcclean2020openfermion} python package (details are given in Methods). Symmetry allows reduction of the problem to the effective Hamiltonian $\hat{\mathcal{H}}_{\mathrm{qubit}}$ which acts on two qubits expressed in the standard Pauli basis ($\mathbb{I}$, $X$, $Y$, and $Z$),
\begin{equation}
\hat{\mathcal{H}}_{\mathrm{qubit}}= \alpha \mathbb{I}\mathbb{I} +\beta Z \mathbb{I}+\gamma \mathbb{I} Z+\delta Z Z +\mu X X
\label{eq:qubithamiltonian}
\end{equation}
with real parameters $\alpha$, $\beta$, $\gamma$, $\delta$ and $\mu$ that depend on the choice of bond length. 
We create ansatz states $\ket {\psi (\vec \theta)}$ made of two path-encoded qubits using single-qubit operations $R(\vec{\theta}_i)$ and an entangling postselected CNOT gate (see Fig.~\ref{fig:applications}.a). The expectation value of $\hat{\mathcal{H}}_{\mathrm{qubit}}$ on $\ket {\psi (\vec \theta)}$ is obtained from the weighted averages of $10 000$ post-processed $2$-photon samples. The classical processor then evaluates a loss function using a gradient-free optimizer based on expectation values obtained from Ascella and corrected with an error mitigation scheme inspired by Ref.~\cite{lee2022error}. Then $\vec \theta$ is updated classically in a feedback loop between Ascella and a classical processor to reach lower and lower energies. Error mitigation helps to consistently reach the ground state energy (see Supplementary~\ref{app:VQE}).
For any initial random parameters and bond length, the algorithm consistently converges to the theoretical eigenvalue within $\pm 0.01$ Hartree in $50$ to $100$ iterations (see Fig.~\ref{fig:applications}.a). The total experiment time per bond length is approximately four times faster than previous photonic VQE experiments of a system with the same number of degrees of freedom \cite{peruzzo2014variational}. In an experiment with fixed initial conditions and bond length, chemical accuracy (an error of $\pm 0.0016$ Hartree) was achieved with a success probability of $93\%$, showing greater accuracy than recent photonic VQE experiments \cite{lee2022error}. These two improvements are due to higher quality single-photon sources and chip control. 

\section*{Photon-native quantum computation} \label{sec:photonic_computation}

%Gate-based quantum computing requires the use of ancilla path modes that increase the physical footprint on the chip as illustrated in Figure~\ref{fig:applications}.a. 
We now demonstrate the operation of Ascella in its native photonic framework,
where the information is directly processed through photonic quantum interferences in an arbitrary unitary transformation and detection.  

\subsection{Photon-based quantum neural network}
We train a quantum neural network~\cite{McClean_2016} on Ascella for a supervised learning classification task. We build a variational quantum algorithm where, taking inspiration from Ref.~\cite{Gan_2022}, we use a native photonic ansatz.  We perform multi-class classification on the well-known IRIS dataset \cite{iris_dataset}. To the best of our knowledge, this is the first experimental implementation of a variational quantum classifier with single photons -- we refer to Ref.~\cite{Havlicek_2019} for a realization on a superconducting platform and to Ref.~\cite{Bartkiewicz_2020} for a $2$-photon classifier based on kernel methods. Following our photon-native approach, we design the ansatz of the variational algorithm directly using the beamsplitters and phase shifters on $5$ modes of Ascella, in which we input $3$ photons. We also implement partial pseudo photon-number resolution by exploiting 4 extra modes of the chip.

We train the model using a see-saw optimization between the chip parameters and the output state parameters that define the measurement observable. Each iteration requires $112$ experiments, one for each data point in the training set, and we gather $50 000$ samples per run. A batch functionality in \textit{Perceval}~\cite{heurtel2023perceval} allows us to send all data points as one job to the server. Details on the ansatz and the training can be found in Methods and Supplementary \ref{app:classification}.
After about $15$ iterations, we find an accuracy of $0.92$ on the training set and $0.95$ on the test set. Fig.~\ref{fig:applications}.b.
provides a summary of the model predictions versus actual labels as a confusion matrix.
 
\subsection{ Boson Sampling with 6 single photons}

Boson Sampling is a sampling problem suited for demonstrating a quantum-over-classical advantage with optical quantum computing platforms \cite{aaronson2011}. The recent demonstrations of quantum advantage~\cite{zhong2020, zhong2021} in the Gaussian Boson Sampling framework \cite{hamilton2017gaussian}  used squeezed light manipulated  in  free-space interferometers to limit optical losses. Genuine single-photon-based Boson Sampling has progressed poorly on integrated chips due to the low efficiency of heralded sources~\cite{zhong2018, paesani2019, gao2019experimental, hoch2021}. Here we demonstrate on-chip Boson Sampling for a record number of $6$ photons with a fully reconfigurable interferometer. A $12 \times 12$ Haar-random unitary matrix is randomly chosen using the dedicated tool in \textit{Perceval}. We record the threshold statistics of all $N$-photon coincidences ($N \in [\![ 1;6 ]\!]$) and acquire in total $340.10^{9}$ samples, with a $6$-photon coincidence rate reduced by the strong bunching of photons in this sampling task down to $0.16$~Hz.

To validate our experimental results, we  discriminate  our collected Boson Sampling statistics from the uniform sampler~\cite{aaronson2013} and distinguishable  sampler~\cite{spagnolo2014} hypotheses (see Fig.~\ref{fig:applications}.c). We also reconstruct the $6$-photon output distribution  from the sampled data and compare it to the ideal output distribution corresponding to the chosen unitary matrix. Both distributions are plotted in Fig.~\ref{fig:applications}.c from which we deduce a fidelity $F=\sum_i\sqrt{p_iq_i}$ and a total variation distance $D=\frac{1}{2}\sum_i|p_i-q_i|$  where
$\{p_i\}$ and $\{q_i\}$ are the ideal and experimental output probability distributions respectively, with $i \in \{1,...,924\}$ labelling the no-collision output configuration of the boson-sampling device~\cite{aaronson2011}. We measure state-of-the-art values  $F=(0.97\pm0.03)$ and  $D=(0.16\pm0.02)$~\cite{wang2018toward, wang2019boson}. Details about the measurement simulation with \textit{Perceval} as well as Boson Sampling with $k$ photons lost ($k\in [\![ 1;4 ]\!]$) are given in Supplementary~\ref{app:boson_sampling}.

\section*{Discussion and resources to scale up}

\begin{figure}[t]
    \centering{
    \includegraphics[width=\linewidth]{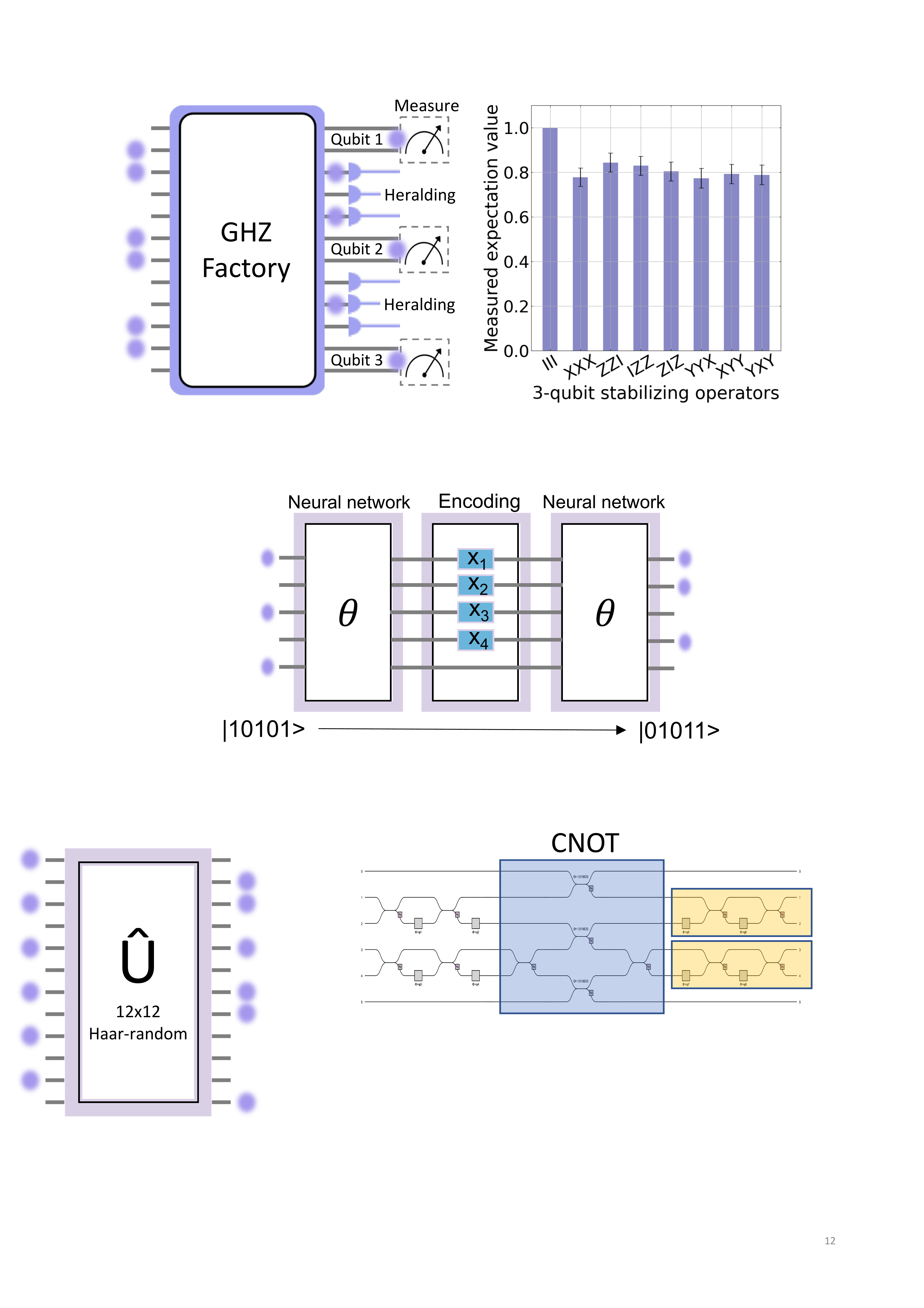}}
    \caption{\textbf{Heralded generation of 3-photon GHZ states.}  Measured expectation values of the stabilizing operators of the heralded 3-photon GHZ state $\ket{\text{GHZ}^+_3}$ yielding a fidelity of ${F_{\text{GHZ}_3^+}}=0.82\pm0.04$.}
    \label{fig:stab_heralded3GHZ}
\end{figure}

The above results demonstrate the suitability of the architecture for near-term quantum computing tasks. The record 4~Hz rate for 6 photons demonstrated here can further be pushed by optimizing each individual component of the platform. This optimization will allow manipulating a large number of photons in a reasonable time (see Supplementary~\ref{app:setup}). Noticeably, the current single-photon source efficiency of $55\%$ at the first lens can be brought to values at least as high as $96\%$ with technology optimization~\cite{wang2020micropillar}. In parallel, the number of modes in the photonic chip can be increased,  the photonic chip side, a record number of $32$ modes was recently reported with a high-transmission glass chip technology~\cite{hoch2021} and progress is continually achieved with silicon nitride-based platforms~\cite{vigliar2021, bao2023}. Our single-photon source technology has demonstrated $\ge$99.5\% indistinguishability~\cite{somaschi2016}, which would bring the two-qubit gates fidelity close to unity~\cite{ralph2002linear}. These improvements will allow linear-optical computing platforms to push to dozens of photons. Scaling-up beyond the limitations of probabilistic linear-optical protocols involves shifting to a measurement-based quantum computing paradigm, which requires the generation of large graph states~\cite{raussendorf2001}. A key step to obtain large graph states is the heralded-generation of entanglement on chip~\cite{li2015}. This is the last feature we implement on Ascella, demonstrating for the first time the heralded generation of $3$-photon GHZ states from a $6$-photon input state.

We use a scheme adapted from Ref.~\cite{li2015,gouriou2019} where $3$ out of the $6$ single photons are detected in $6$ optical modes  identified as $8$ heralding states (see Methods). Four of them herald the generation of the state $\ket{\text{GHZ}^+_3}=(\ket{000}+\ket{111})/\sqrt{2}$. The  fidelity of the heralded state to the target state is  characterized on Ascella using ${F_{\text{GHZ}_3^+}}=\frac{1}{8}\sum_i \langle S_i \rangle $, where $S_i \in$ $\{III, XXX, ZZI, IZZ, ZIZ, -YYX, -XYY, -YXY\}$ are the stabilizing operators of the target state and are experimentally accessed through the $3$-qubit operators $XXX$, $ZZZ$, $YYX$, $XYY$, and $YXY$. All expectation values $\langle S_i \rangle$ are reported in Fig.~\ref{fig:stab_heralded3GHZ}, and yield a fidelity of ${F_{\text{GHZ}_3^+}}=0.82\pm0.04$ (see Methods) providing the first benchmarking of such heralded state generation. 

Such heralded entanglement schemes combined with the recent demonstration of efficient generation of linear cluster states directly from the same quantum dot source technology~\cite{coste2022high} open the path to fault tolerant quantum computing with reasonable hardware resource overheads. 

\section*{Acknowledgements}
The authors would like to thank A.\ White for fruitful feedback, R.\ Osellame and his team for valuable interactions on the generation of heralded GHZ states, JJ.\ Dormard, G.\ Parent, J.\ Herlent for support in the electronic modules.
The authors acknowledge I.\ Maillette de Buy Wenniger, V.\ Guichard, F.\ Hoch, and A.\ Henry for preliminary work on the classification experiment.
This work has received funding from the European Union’s Horizon 2020 Research and Innovation Programme QUDOT-TECH under the Marie Sklodowska Curie Grant Agreement No.\ 861097, and from BPI France Concours Innovation PIA3 projects DOS0148634/00 and DOS0148633/00 – Reconfigurable Optical Quantum Computing.\\

\section*{Author contributions}
Correspondance should be addressed to J. Senellart ({\tt jean.senellart@quandela.com}) and N. Somaschi ({\tt niccolo.somaschi@quandela.com}).

The QD single photon source was fabricated by A.P., N.\ Marg., W.H., S.B.\ and H.A.;
its optical characterization was performed by P. St.;
A.F., E.I., M.P., M.B., O.A.\ and A.B.\ integrated all the hardware components under the close supervision of N.M.;
A.F., N.M., N.B.\ and J.S.\ developed the machine-learned chip control;
E.I., A.F.\ and M.P.\ developed the software layer controlling all instruments and realizing the remote tasks; 
M.V.\ architected the software stack, E.B.\ implemented the control code within \textit{Perceval} and A.B.\ wrote the interface between \textit{Perceval} and the cloud worker;
R.M., S.W.\, P.Si.\ and J.S.\ conducted the gate benchmarking;
M.P.\ and R.M.\ ran the Boson Sampling task; M.P.\ implemented the 6-photon indistinguishability test and heralded 3-photon GHZ state generation;
A.S. and D.F. realized the classification experiment, P.Si.\ and P-E.E. defined and realized the VQE;
S.M.\ supervised and coordinated the theoretical work;
P.Se.\ and N.S.\ guided the source fabrication process and hardware integration; J.S.\ coordinated the full assembly of the hardware and software;
N.S.\ supervised the overall project;
M.P., P-E.E., A.S., R.M., N.M., P.Si., A.F., S.W., N.S., N.B, S.M., J.S., P.Se.\ wrote the paper.

\section*{Data \& code availability}

The data generated as part of this work is available upon reasonable request from the corresponding authors. The code used to run the presented applications is available at {\tt https://github.com/Quandela/Ascella}.

%%-- Bibliography
\typeout{}
\hypersetup{breaklinks=true}
% use only in methods
 \nocite{hein2006}

\bibliographystyle{unsrtnat}
\bibliography{final_biblio.bib}

\section*{Methods}
\subsection*{Architecture}

Ascella is accessible remotely via a cloud service~\cite{QuandelaCloud}. Tasks can be dispatched either to Ascella, to a perfect simulator or to a noisy simulator through a generic scheduler handling user access limitations and task prioritization. Following a compilation and transpilation process, Ascella then sets the demultiplexer configuration and the photonic circuit phases to apply  the required unitary matrix to the input state. For applications like quantum machine learning (QML) for which each training data sample corresponds to a task, users can prepare and send a batch of tasks that will execute sequentially on the QPU with fast incremental chip reconfiguration and without any communication overhead. 

\subsection*{Chip control benchmarking}

We benchmark the transpilation process by applying $300$ random phase configurations on the photonic chip and measuring the photon countrates at the $12$ outputs. We compare them to a simulation of the chip which is takes into account the estimated directional coupler reflectivities and relative output losses (see Supplementary~\ref{app:chip_charac} for values).  We quantify the difference between the measured and simulated values using the total variation distance (TVD). At $925$ nm, with a standard characterization of the chip based on interference fringes measurements~\cite{taballione2020}, the TVD evaluated on the configurations is ($21 \pm 11$) \%, where the error bar is the standard deviation of the dataset. At the operating wavelength of our single-photon source (928 nm), with our machine learning process, we achieve a TVD of (3.0 ± 1.3) \%, greatly improving our control over the chip. The relative variation on the obtained average TVD between successive benchmarkings is of the order of $3$ \%, showing repeatability of the obtained value.

\subsection*{Variational Quantum Eigensolver} \label{metho:VQE}

The ansatz for the VQE algorithm implements the gate-based circuit shown in Fig.~\ref{fig:applications}.a which consists of a generic $2$-qubit state generator. It comprises single-qubit rotations and a CNOT gate~\cite{ralph2002linear}. This is implemented on $6$ modes (modes $1$ to $6$) comprising two path-encoded qubits and two extra modes for the postselected Ralph CNOT. 
Arbitrary rotations are implemented via tunable Mach-Zehnder interferometers with thermo-optic phase shifters. Extra phase shifters are used to mitigate systematic errors in the reflectivity of beamsplitters and to converge faster to the ground state energy.

\subsection*{Boson Sampling}

Two statistical tests are used to discriminate the experimental data against the uniform sampler and distinguishable particle hypotheses. The A \& A counter and the likelihood ratio counter, respectively, are increased or decreased according to a likelihood ratio test. The A \& A counter $A$ is defined as~\cite{aaronson2013, hoch2021}
\begin{equation*}
A_k:=~\left\{ \begin{aligned} 
A_{k-1}+1~\text{if}~\mathcal{P}\ge \left(\frac{n}{m}\right)^2\\
A_{k-1}-1~\text{if}~\mathcal{P} < \left(\frac{n}{m}\right)^2\end{aligned}
\right.
\end{equation*}
\noindent where $n$ and $m$ are, respectively, the number of photons and optical modes, and $\mathcal{P}:=\prod_i\sum_j|U_{ij}|^2$, where $i$ labels the modes in which photons are detected, $j$ the input modes and $U$ is the unitary sampling matrix.

The likelihood ratio counter $C$ is defined as~\cite{spagnolo2014, hoch2021}
\begin{equation*}
C_k:=~\left\{ \begin{aligned} 
C_{k-1}+1~\text{if}~\mathcal{L}\ge \left(\frac{n}{m}\right)^2\\
C_{k-1}-1~\text{if}~\mathcal{L} < \left(\frac{n}{m}\right)^2\end{aligned}
\right.
\end{equation*}
\noindent where $\mathcal{L}:=\frac{q}{p}$ with $q:=|\text{Perm}(U_{(ij)})|^2$, $p:=|\text{Perm}(|U_{(ij)}|^2)$ and $U_{(ij)}$ denoting the sub-matrix restricted to the input labels $i$ and output labels $j$.

\subsection*{Photon-based quantum neural network} \label{metho:classification}

We build the ansatz of our variational quantum classifier using modes $3$ to $7$ of Ascella. We input three photons into the chip, in modes $3$, $5$ and $7$. We use $32$ of the reconfigurable thermo-optic phase shifters as the variational parameters, and $4$ phase shifters in the middle of the chip for the data encoding. We use extra modes for pseudo photon-number resolution (PNR): by setting four phase shifters to $\pi/2$ in the final layer of the chip, we redirect a portion of the photons from modes $3$ and $7$ into modes $1$, $2$ and $8$, $9$ respectively. For the classical optimization process, we use a see-saw approach based on Gaussian processes and Nelder-Mead optimizers. More details regarding the circuit ansatz, model definition, pseudo PNR, and the optimization methods are in Supplementary \ref{app:classification}.

\subsection*{Heralded three-photon GHZ on-chip generation} \label{metho:3GHZ}

The generation of a path-encoded $3$-photon GHZ state is characterized with three reconfigurable integrated Mach-Zehnder interferometers (MZI$_i$, $i=1, 2, 3$). The layout of the optical circuit is provided in the Supplementary~\ref{app:ghz}. The output state of the circuit is given by~\cite{gouriou2019} 
\begin{equation*}
    \begin{split}
    \ket{\text{Out}} & =\frac{1}{16} (-\ket{\text{GHZ}^-_3}\ket{h_1} +\ket{\text{GHZ}^-_3} \left[\ket{h_4} + \ket{h_6}+ \ket{h_7} \right] \\ 
    & - i \ket{\text{GHZ}^+_3}\ket{h_8} + i \ket{\text{GHZ}^+_3} \left[\ket{h_2} + \ket{h_3}+ \ket{h_5} \right]).
    \end{split}
\end{equation*}
\noindent We target the state $\ket{\text{GHZ}^+_3}$, where $\ket{\text{GHZ}^{\pm}_3}=(\ket{000}\pm\ket{111})/\sqrt{2}$, which is obtained by conditioning the analysis of the state on the detection of one of the heralding states $\ket{h_2}$, $\ket{h_3}$, $\ket{h_5}$, and $\ket{h_8}$.

The heralding channels signal the generation of a specific GHZ state. The heralding conditions for the generation of $\ket{\text{GHZ}^+_3}$ are
\begin{equation*}
    \left\{ \begin{aligned} 
    \ket{h_2}=\ket{0_2 1_3 0_4 1_7 1_8 0_9}\\
    \ket{h_3}=\ket{1_2 0_3 0_4 1_7 0_8 1_9}\\
    \ket{h_5}=\ket{1_2 0_3 1_4 0_7 1_8 0_9}\\
    \ket{h_8}=\ket{0_2 1_3 1_4 0_7 0_8 1_9}
\end{aligned} \right.
\end{equation*}
\noindent where $\ket{0_i}~(\ket{1_i})$ correspond to detecting $0$ ($1$) photons in mode $i$ (modes are labelled from $1$ to $12$ from top to bottom). %The heralding conditions for the generation of $\ket{\text{GHZ}^-_3}=(\ket{000}-\ket{111})/\sqrt{2}$ are

The state $\ket{\text{GHZ}^+_3}$  is a stabiliser state, and therefore can uniquely be expressed in terms of its stabilisers~\cite{hein2006}.

\begin{equation}
\ket{\text{GHZ}^+_3}\bra{\text{GHZ}^+_3}=\sum_{S_i \in \mathcal{S}}\frac{1}{|\mathcal{S}|}S_i,
\end{equation}

\noindent where $S_i$ is a stabiliser of $\ket{\text{GHZ}^+_3}$, $\mathcal{S}$
is the stabiliser group of $\ket{\text{GHZ}^+_3}$, and $|\mathcal{S}|$ is the number of elements of $\mathcal{S}$.
The fidelity of some experimental implementation $\rho$ of $\ket{\text{GHZ}^+_3}$ is given by
$$F_{\text{GHZ}_3^+}=\text{Tr}(\ket{\text{GHZ}^+_3}\bra{\text{GHZ}^+_3}\rho),$$
Plugging the expansion of $\ket{\text{GHZ}^+_3}\bra{\text{GHZ}^+_3}$ into $F_{\text{GHZ}_3^+}$ and using linearity of the trace, one obtains
$$F_{\text{GHZ}_3^+}=\frac{1}{|\mathcal{S}|}\sum_{S_i \in \mathcal{S}}\mathsf{Trace}(S_i \rho)=\frac{1}{|\mathcal{S}|}\sum_{S_i \in \mathcal{S}}<S_i>.$$

% Supplementary Information
\clearpage
\pagebreak
\widetext
\begin{center}
    \textbf{\large Supplementary Information}
\end{center}

\setcounter{section}{0}
\setcounter{equation}{0}
\setcounter{figure}{0}
\setcounter{table}{0}
\setcounter{page}{1}
\makeatletter
\renewcommand{\thetable}{S\arabic{table}}
\renewcommand{\thesection}{S-\Roman{section}}
\renewcommand{\theequation}{S\arabic{equation}}
\renewcommand{\thefigure}{S\arabic{figure}}

%!TEX root =  main.tex

\section{Single-photon source} \label{app:SPS}

The single-photon source is based on a gated InGaAs quantum dot (QD) embedded in a monolithic micropillar cavity~\cite{somaschi2016} cooled down to $5$K. It is optically excited using the near-resonant LA-phonon-assisted excitation scheme~\cite{thomas2021} at an \qty{80}{\mega\hertz} rate. The neutral QD emits linearly-polarized single photons at \qty{928}{\nano\meter} with a lifetime of \qty{92}{\pico\second} and a $55\%$ first lens brightness. After spectrally filtering the remaining excitation laser using free-space spectral bandpass filters (FWHM=$800$~pm) and coupling into a single mode fiber, the single-photon source device shows a low multiphoton component with a single-photon purity $\mathcal{P}= 1 - g^{(2)}(0) > \SI{99}{\%}$ and a $2$-photon indistinguishability $M_s>94\,\%$.

\begin{figure}[h]
\begin{tabular}{cc}
    \begin{minipage}{0.5\columnwidth}
            \centering
        \includegraphics[width=\textwidth]{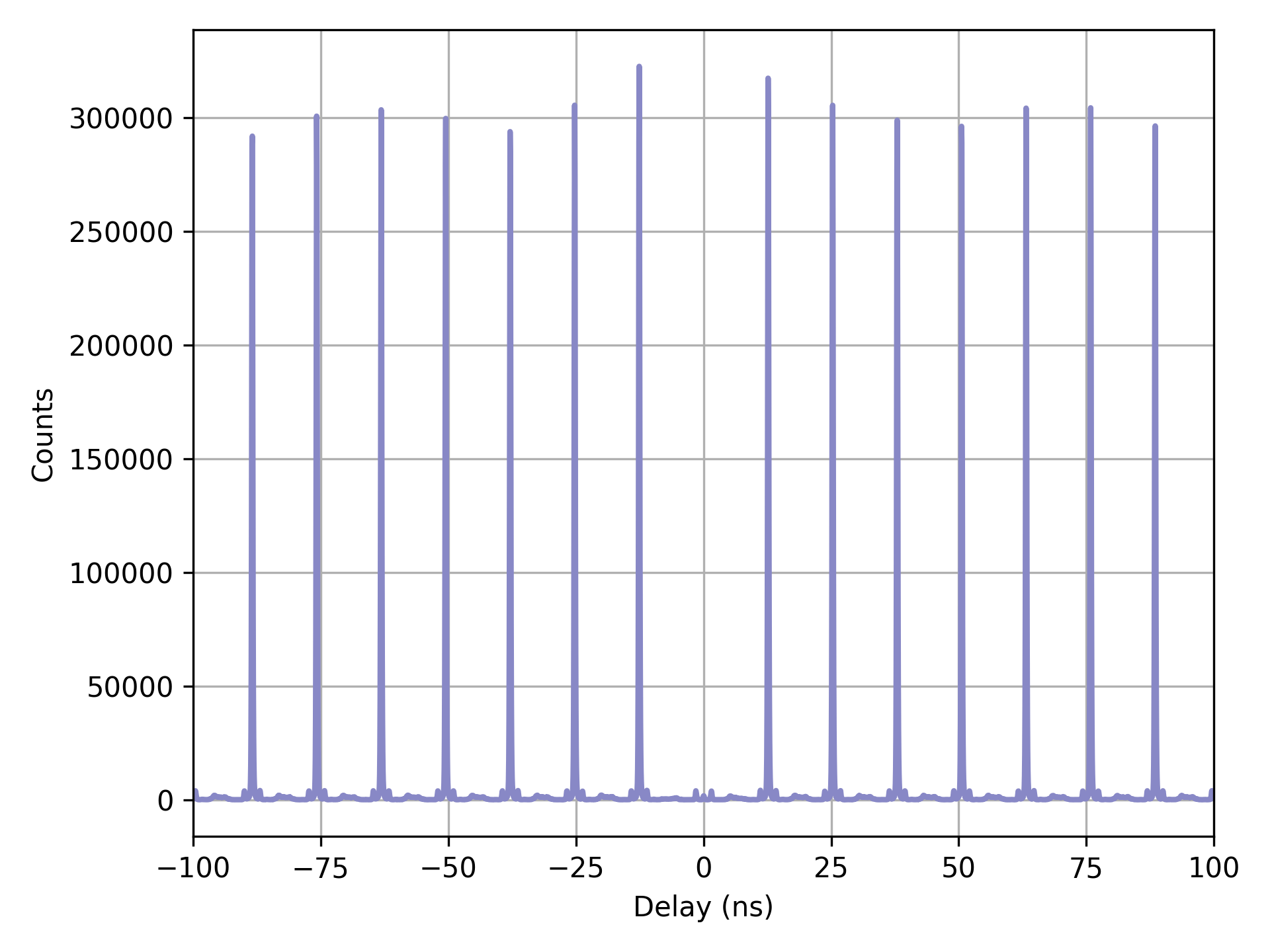}
       % \caption{}
        %\label{subfig:g2}
\end{minipage}
            &       
      \begin{minipage}{0.5\columnwidth}
             \centering
        \includegraphics[width=\textwidth]{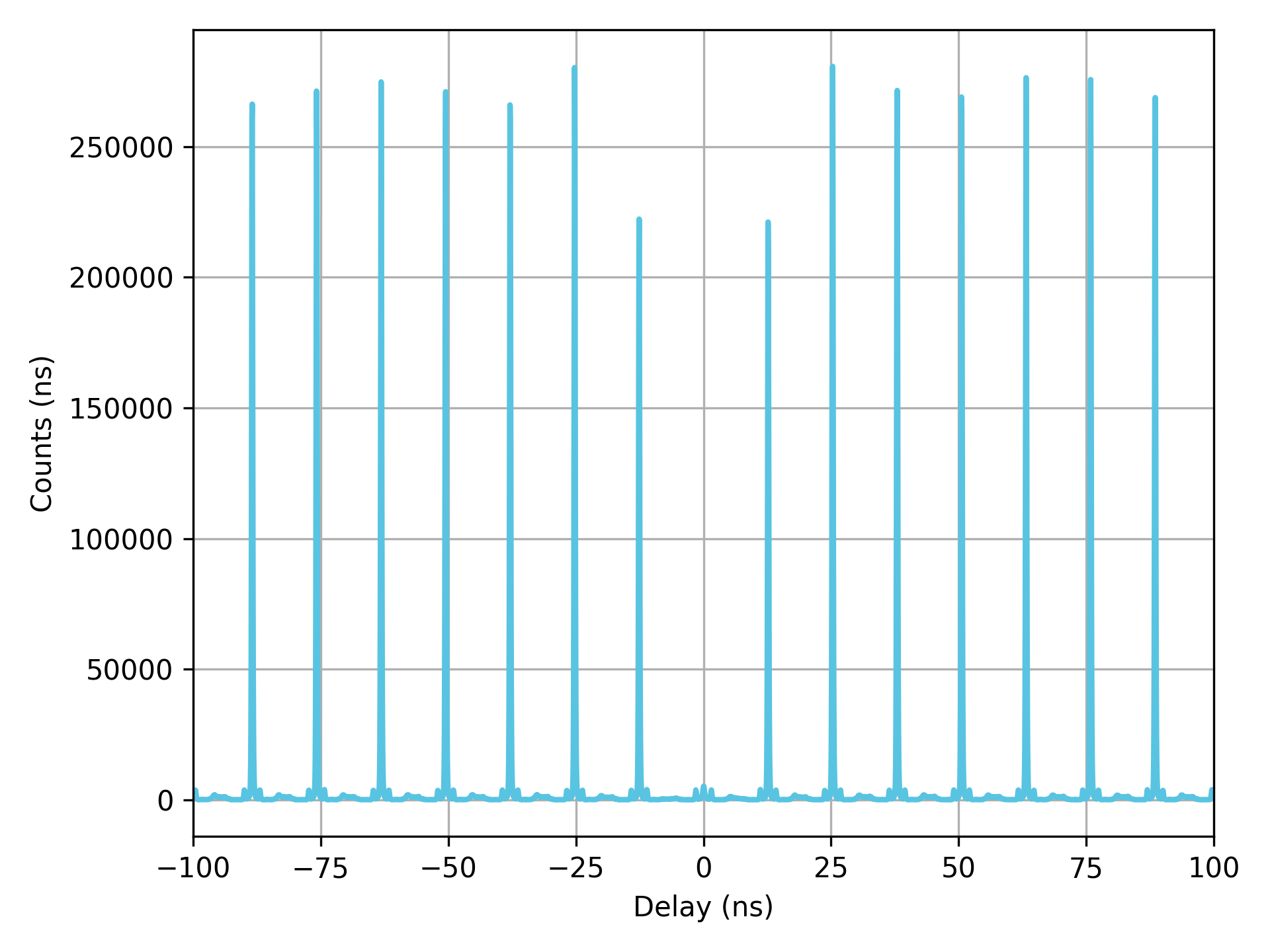}
        %\caption{}
        %\label{subfig:hom}
\end{minipage}      
 \end{tabular}
   \caption{
    \textbf{The single-photon source delivers pure and indistinguishable single photons.} 
    Single-photon purity is quantified by the normalized second-order correlation function $g^{(2)}(0) = (7.32 \pm 0.07) \times 10^{-3}$ (left) and the Hong-Ou-Mandel visibility $V_\text{HOM} = 0.9296 \pm 0.0003$ (right) yielding a corrected  $2$-photon indistinguishability of $M_s=0.9438\pm0.0003$~\cite{ollivier2021}. The histograms were integrated for \qty{5}{\second} and the peak integration window is \qty{1}{\nano\second}. The integrated values are obtained with no background subtraction.
    }
    \label{fig:source_values}
\end{figure}

\section{Optical setup} \label{app:setup}

The overall transmission of the optical setup (see Fig.~\ref{fig:main_hardware}.a) is characterized by the transmission or efficiency of each module. A precise loss budget of the optical setup is provided in Tab.~\ref{tab:loss}.

\begin{table}[h]
    \begin{tabular}{ ||c|c|c|c|| } 
    \hline
    Module & Transmission/Efficiency & Near-term targets\\
    \hline
    First lens brightness & 55$\,\%$ & 80\%~\cite{gazzano2013bright}\\ 
    Single-mode fiber coupling & 70$\,\%$ & 85\%~\cite{snijders2018fiber}\\
    Spectral Filtering module & 75$\,\%$ & $>$82\%[$^*$]\\
    Demultiplexer & 70$\,\%$ & $>$80\%[$^*$]  \\
    PIC insertion and transmission & 45$\,\%$ & 70\%~\cite{della2008}\\
    SNSPDs & 92$\,\%$ & $>$95\%[$^{**}$]\\ 
    \hline
    \textbf{Total} & \textbf{$8.4\pm0.2\,\%$}  & 27\%\\ 
    \hline \hline
    \textbf{Pump laser repetition rate} & \qty{80}{\mega\hertz} & \qty{320}{\mega\hertz}~\cite{anton2019}\\ 
    \hline
    \textbf{6-photon countrate} & \qty{4}{\hertz} & $\sim$\qty{35}{\kilo\hertz} (computed)\\ 
    \hline
    \textbf{12-photon countrate} & \qty{200}{\nano\hertz} (computed) & $\sim$\qty{10}{\hertz} (computed)\\
    \hline \hline
    \end{tabular}
    \caption{Current loss budget of the optical setup (Transmission/Efficiency) and near-term targets. [$^*$] Optical module in development at Quandela. [$^{**}$] Commercially available products}\label{tab:loss}
\end{table}

%DMX
Because the photon input ports on the chip are maintained fixed, the transpilation process also uses the chip's universality to input arbitrary photon configurations by implementing in reality unitary matrices of the type $\hat{U}\times\hat{U}_\text{perm}$, where $\hat{U}$ is the initial unitary matrix and $\hat{U}_\text{perm}$ a permutation matrix yielding the required photon input configuration. If less than $6$ photons are needed, mechanical shutters block the paths of additional photons at the demultiplexing stage.

% Detection
Ascella's detection module is composed of superconducting nanowire single-photon detectors showing an average detection efficiency of $92\,\%$ and dark countrates under $20$ Hz. Detection events are digitalized using a time-to-digital converter (Swabian instruments) and post-processed for sampling $N$-photon ($1\leq N\leq6$) coincidences between all $12$ detectors within a \qty{1}{\nano\second} coincidence window. 

% Monitoring and optimization 
%Several metrics are measured and logged such as total transmissions through 12 different paths of the photonic circuit, on-chip single photon purity and indistinguishability, detectors dark counts, 2- to 6-photon coincidence rates, demultiplexer state preparation, laser excitation power, cryostats temperatures, and ambient temperature. 
The polarization in the delay fibers is optimized automatically to ensure a maximal transmission to the polarization-selective photonic circuit. The excitation laser power is similarly stabilized, ensuring optimal brightness and photon purity. 

\section{Multiphoton interference characterization} \label{app:indistinguishability}

In this section, we measure all pairwise $2$-photon indistinguishabilities, and the \textit{genuine} $4$- and $6$-photon indistinguishability of the input state of Ascella, i.e. the probability that the \textit{n} ($n=4$ or $n=6$) photons are identical~\cite{brod2019, pont2022quantifying}. This initial characterization of our input state sets the basis for future practical applications on our platform. It also allows us to fine-tune our simulator to be able to reproduce experimental results with a good agreement. 

We first measure the $2$-photon pairwise indistinguishability $M_{ij}$ ($i,j=1,...,6$) between all $C_{6}^2=15$ photon pairs. The reconfigurable chip is set to successively connect each pair with a Mach-Zehnder interferometer (MZI). We vary the internal phase of the  MZI to measure the correlated $\phi=\pi/2$ (uncorrelated $\phi=\pi$) $2$-photon coincidences at zero time delay, which gives access to the visibility of the $2$-photon interference fringe. The imperfect single-photon purity~\cite{ollivier2021} of our QD-source ($g^{(2)}(0)=0.0075$), is taken into account to compute $M_{ij}$ for all $15$ pairs. The values of $M_{ij}$ for each of the $15$ pairs is reported in the \textit{indistinguishability matrix} $\mathcal{M}$ where $\mathcal{M}_{ij}=M_{ij}$ and $\mathcal{M}_{ii}=1$.

\begin{equation*}
    \mathcal{M} = \begin{bmatrix}
            1 & 0.942 & 0.921 & 0.924 & 0.917 & 0.914\\
            & \ddots & 0.935 & 0.925 & 0.924 & 0.919\\
            & & \ddots & 0.932 & 0.911 & 0.925\\
            & & & \ddots & 0.943 & 0.941\\
            &  & & & \ddots & 0.942\\
            & & & &  & 1
    \end{bmatrix}
\end{equation*}

\begin{figure}[h]
    \centering
    \includegraphics[width=0.6\linewidth]{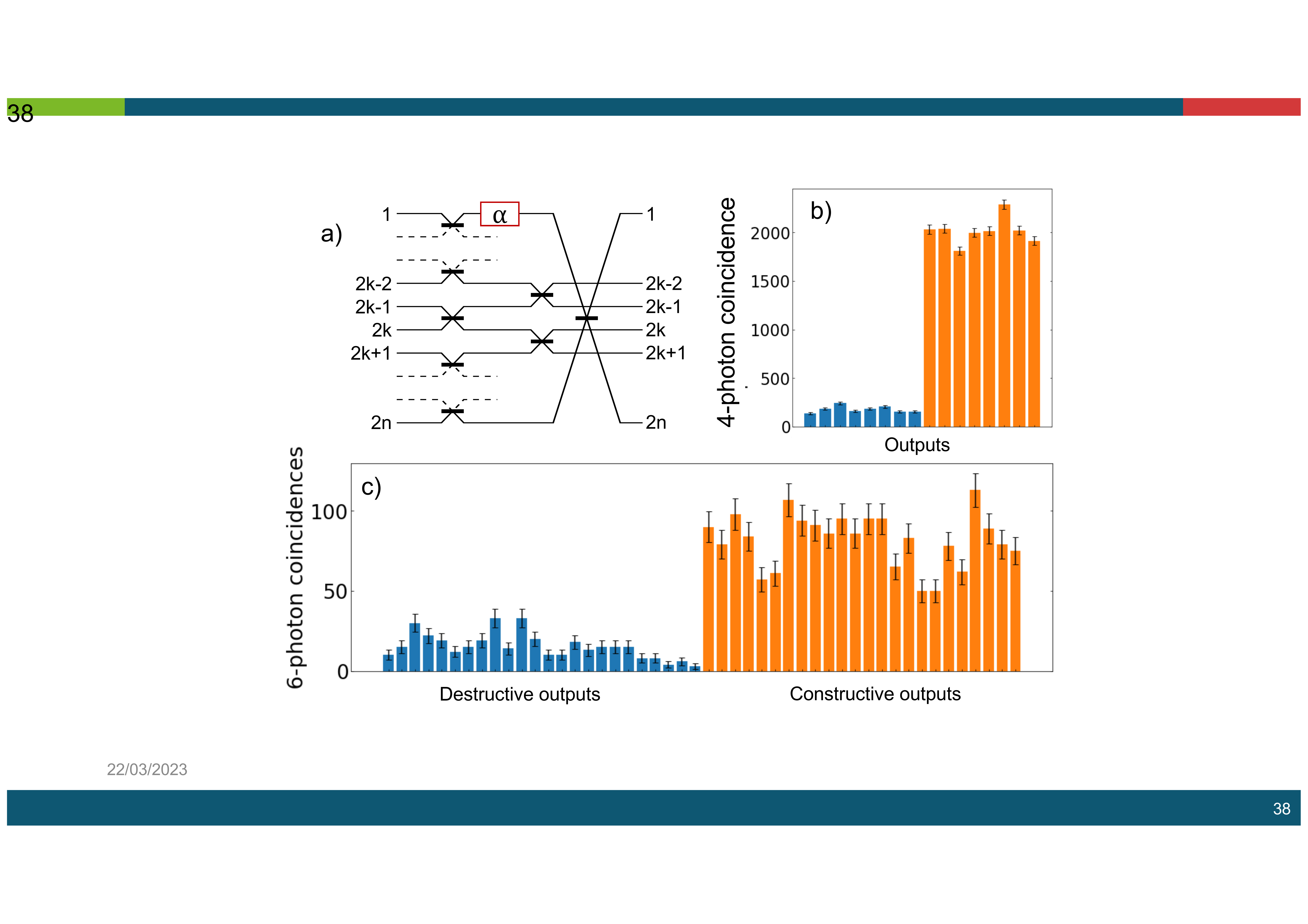}
    \caption{Genuine indistinguishability of the input multiphoton state. (a) General layout of the multiport interferometer used to measure the probability that the $N$ photons are identical. (b-c) Histogram of $4$- ($6$-) photon outputs that undergo destructive (blue) and constructive (orange) interferences in an $8$- ($12$-) mode version of the interferometer presented in (a) fed with $4$ ($6$) photons. In (b) the genuine $4$-photon indistinguishability is $p_{4}=0.85 \pm 0.02$. In (c) the genuine $6$-photon indistinguishability is $p_{6}=0.76 \pm 0.02$.}
    \label{fig:n_indistinguishability}
\end{figure}

The genuine $N$-photon indistinguishability is the probability $p_{N}$ that all $N$ photons are identical. To quantify experimentally the genuine $4$- and $6$-photon indistinguishability of our input multiphoton state we implement on the reconfigurable QPU $8$-mode and $12$-mode versions of the cyclic interferometer (first introduced in~\cite{pont2022quantifying}) whose general layout is presented in Fig.~\ref{fig:n_indistinguishability}.a. First, the single internal phase of the interferometer $\alpha$ is set to $0$~($2\pi$). Each odd input port of the interferometer is fed with single photons. We detect the output states corresponding to one photon per pair of output ports ($2k$, $2k+1$) (see Fig.~\ref{fig:n_indistinguishability}.a). In Fig.~\ref{fig:n_indistinguishability}.b-c we present the experimental output distribution for all outputs corresponding to constructive (orange) and destructive (blue) $n$-photon interference ($N=4$ for b. and $N=6$ for c.). The visibility of the interference fringe is the genuine $N$-photon indistinguishability. We experimentally measure $p_\text{{4}}=0.85 \pm 0.002$ $4$-photon indistinguishability for photons $\{1, 2, 3, 4\}$ and $p_\text{{6}}=0.76 \pm 0.02$ $6$-photon indistinguishability. This work constitutes the first experimental realization of this protocol for $6$ photons, and sets a new state-of-the-art for genuine $4$- and $6$-photon indistinguishability. 

To further study our ability to drive the internal phase of the interferometer shown in Fig.~\ref{fig:n_indistinguishability}.a, we scan the single internal phase $\alpha$ to measure the full interference fringe visibility for $4$-photon interferences with photons $\{1, 2, 3, 4\}$. Note that for each value of the internal phase $\alpha$ we compute the associated unitary matrix using \textit{Perceval} and transpile the circuit to be implemented on the QPU. After normalization of the $4$-photon counts, we fit the theoretical interference fringe~\cite{pont2022quantifying} $p_4=1 \pm c_1\cos(a\cdot\alpha+b)$. In the ideal case we expect $a_{\text{ideal}}=1$ and $b_{\text{ideal}}=0$. The experimental data is well fitted with $a=1.00\pm0.01$ and $b=-0.06\pm0.08$~rad, which shows that the transpilation can very accurately implement an $8 \times 8$ unitary matrix. 

\begin{figure}[h]
    \centering
    \includegraphics[width=0.5\linewidth]{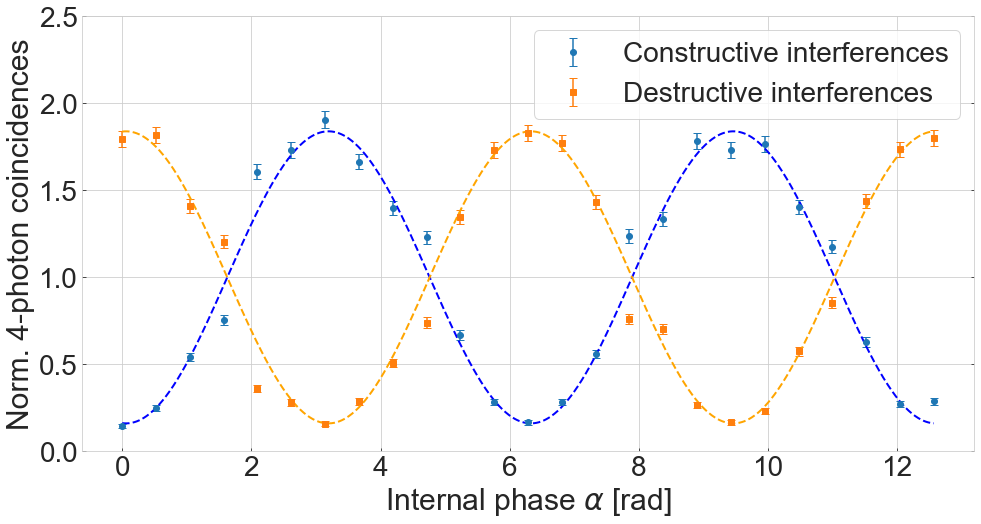}
    \caption{Total normalized four-photon coincidence rate (sum of all eight output states) for the constructive and destructive outputs, as a function of the internal phase $\alpha$. The error bars are computed assuming a shot noise limited error on the detected $4$-fold coincidences}
    \label{fig:4-photon_fringe}
\end{figure}

\section{Photonic chip characterization}
\label{app:chip_charac}

\begin{figure}[h]
\begin{tabular}{cc}
    \begin{minipage}{0.5\columnwidth}
        \centering
        \includegraphics[width=\textwidth]{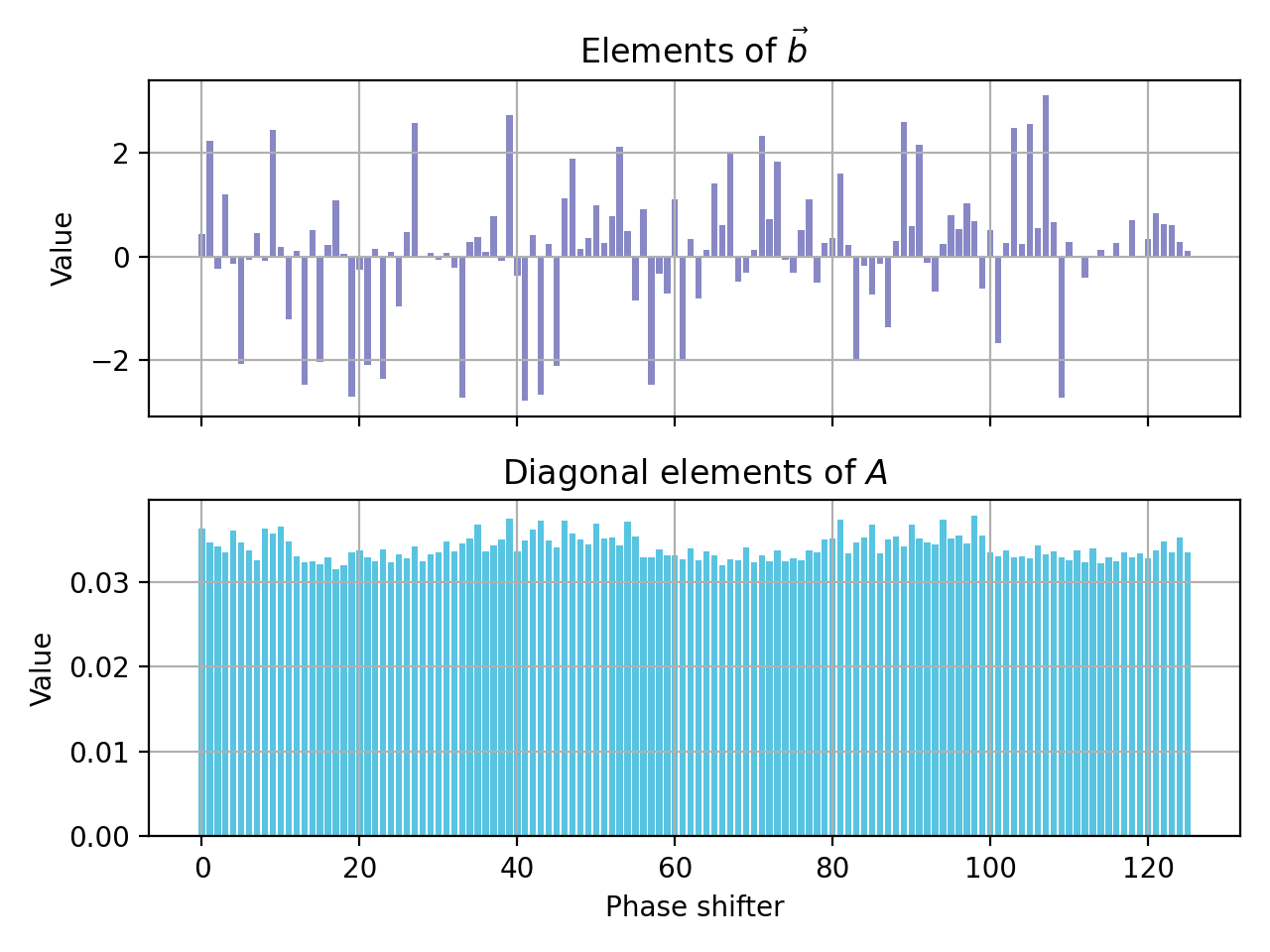}
        \caption{}
        \label{subfig:c0}
    \end{minipage}
    &       
    \begin{minipage}{0.5\columnwidth}
        \centering
        \includegraphics[width=\textwidth]{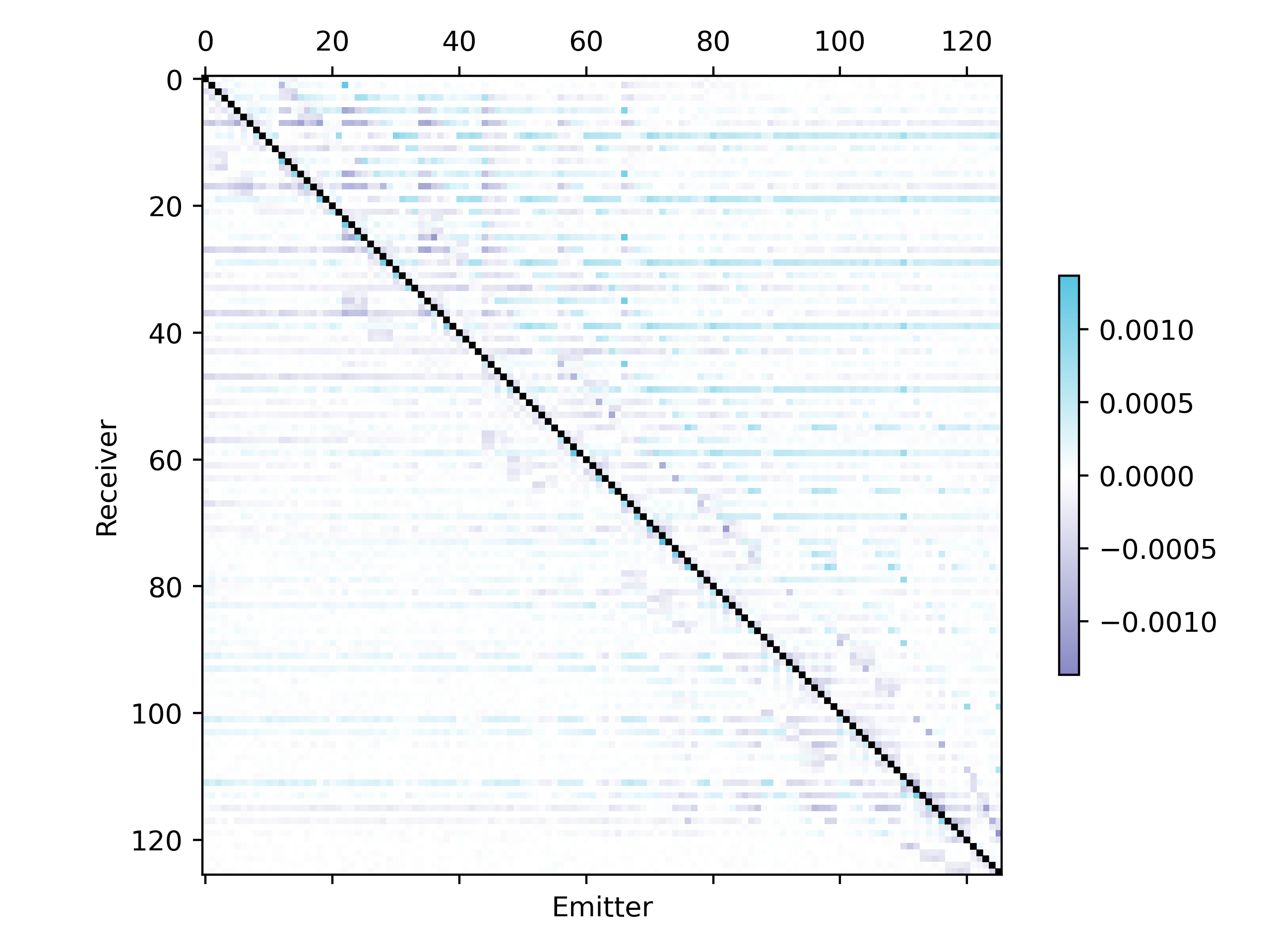}
        \caption{}
        \label{subfig:c2}
        \end{minipage}      
    \\
    \begin{minipage}{0.5\columnwidth}
        \centering
        \includegraphics[width=\textwidth]{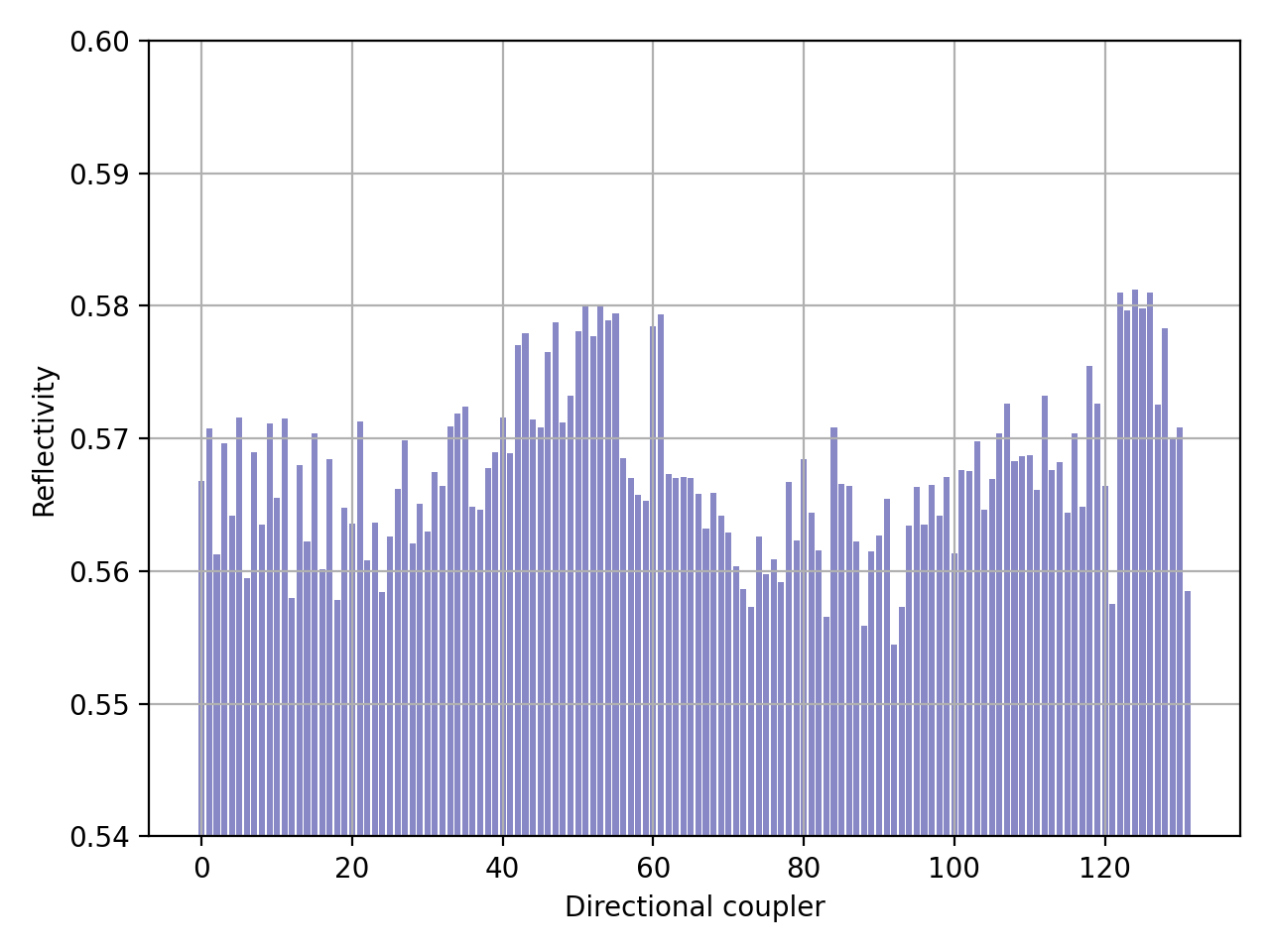}
        \caption{}
        \label{subfig:refl}
    \end{minipage}
    &       
    \begin{minipage}{0.5\columnwidth}
        \centering
        \includegraphics[width=\textwidth]{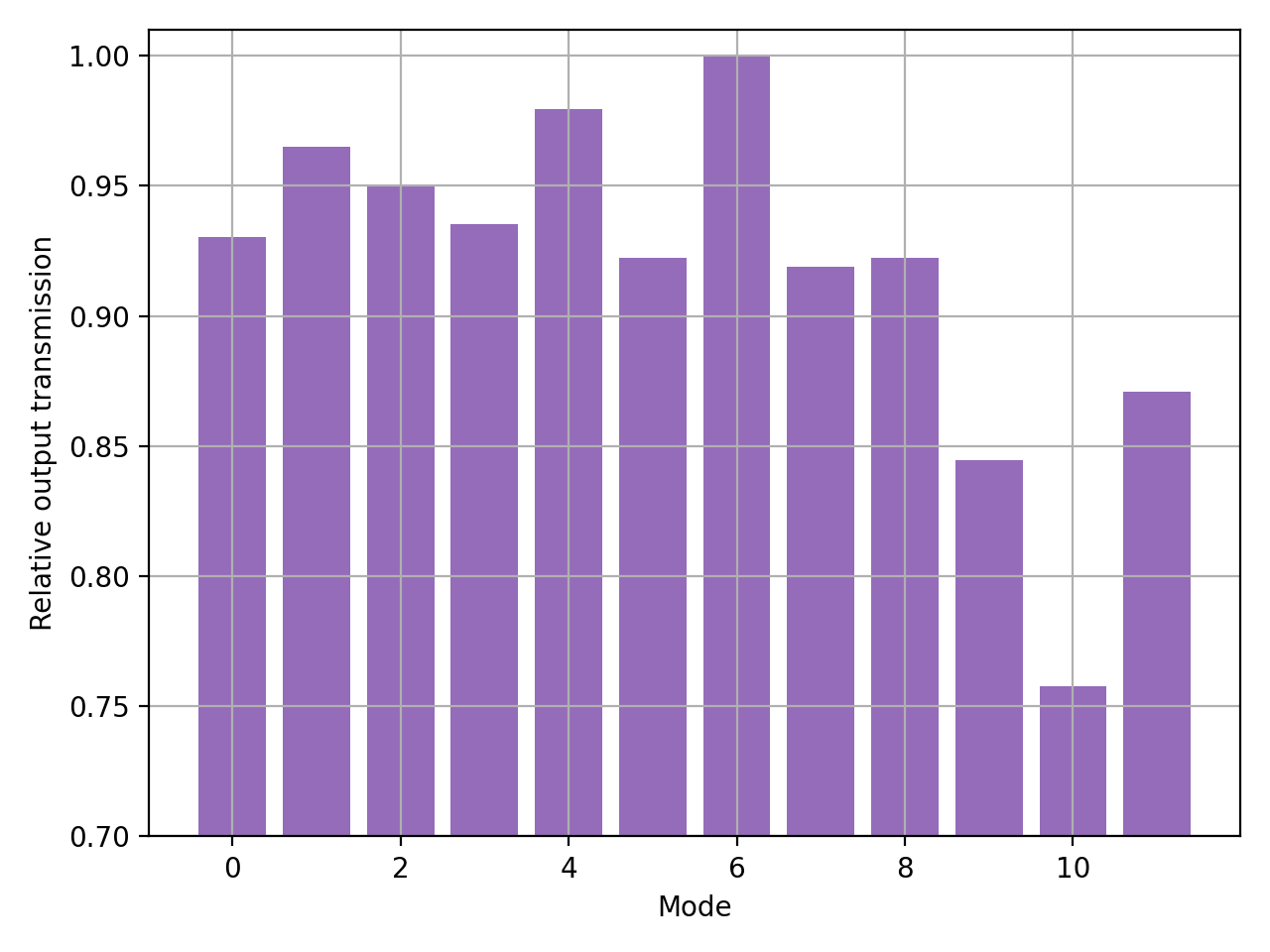}
        \caption{}
        \label{subfig:losses}
    \end{minipage}      
\end{tabular}
\caption{ 
\textbf{Large-scale photonic chip imperfections estimated by a machine-learning process.} 
The phase shifter phase-voltage relation is modelled by a matrix relation of the form $\vec{\phi} = A\vec{V}^{\odot2}+\vec{b}$, where $\vec{\phi}$ and $\vec{V}^{\odot2}$ are vectors containing respectively the applied phase shifts and squared voltages. We show on \ref{subfig:c0} the diagonal elements of $A$ and the elements of $\vec{b}$, and on \ref{subfig:c2} we show the off-diagonal elements of $A$, which account for thermal crosstalk. \ref{subfig:refl} represents the values of the reflectivity of the on-chip directional couplers. \ref{subfig:losses} displays the relative output losses per mode, scaled such that the maximum value is equal to $1$.
}
\label{fig:charac_values}
\end{figure}

We use a machine learning-based process to characterize the photonic chip. The $126$ on-chip thermo-optic phase shifters generate heat via the Joule effect, thus they can be modelled by a relation of the form $\vec{\phi} = A\vec{V}^{\odot2}+\vec{b}$ between the vector $\vec{\phi}$ containing all $126$ physically implemented phases and the vector $\vec{V}$ corresponding to the $126$ applied voltages squared. $^{\odot2}$ represents element-wise squaring. Off-diagonal elements of the $126\times 126$ matrix $A$ represent thermal crosstalk between phase shifters. A process based on machine-learning techniques optimizes the values of $A$ and $\vec{b}$, which represent $\approx 16 000$ free parameters to determine. The same process also estimates individual directional coupler reflectivities and relative output losses. We show on Fig. \ref{fig:charac_values} the estimated values. The elements of $\vec{b}$ have values $\SI{0.1(12)}{rad}$, and the diagonal elements of $A$ have values $\SI{0.034(1)}{rad\per V^2}$, ensuring that on average a $\pi$-phase shift can be achieved by applying around $10$~V on a phase shifter. The matrix $A$ seems to show long-range interactions between phase shifters, but these are, in reality, artefacts arising from certain transformations on $A$ that leave the output quantum state unchanged. The directional coupler reflectivities have values $\SI{56.8(6)}{\%}$. One can observe regions of low and high reflectivities on Fig.~\ref{subfig:refl}, which is a signature of the photonic chip's folding; that is the interferometer is not laid out in a straight manner, but is folded to increase compacity. For the output losses, we notice that mode $10$ has a a relative transmission of $75 \%$ compared to mode $6$. This result was confirmed on two separate detection systems (power meter array and single-photon detectors), hinting that the defect lies in the photonic chip and not in the photon detectors.

\section{Benchmarking logic gates}
\label{app:benchmarking}

In this supplementary section, we outline the method of Ref.~\cite{MW23} applied to benchmark $1$-, $2$-, and $3$- qubit gates implemented by Ascella. We then give an explicit example of the method by deriving $F_\text{avg}$ for the $T$-gate. Finally, we describe the general approach taken to obtain $F_\text{avg}$ for multi-qubit gates and discuss the application to the $2$- and $3$-qubit gates benchmarked in the main text.

\subsection{Symmetry-based benchmarking}
Let $\ket{\boldsymbol{i}}:=\ket{i_1, \dots ,i_n }$ with $i_j \in \{0,1\}$ for $j \in \{1, \dots, n\}$ denote an $n$-qubit computational basis state. A noisy implementation of the gate unitary $U$ is given by $\Lambda \circ \mathbf{U}$(.), which is a completely positive trace preserving (CPTP) map \cite{nielsen2002quantum} acting on $n$-qubit density matrices $\rho$ as $\Lambda \circ \mathbf{U} (\rho)=\Lambda(U\rho U^{\dagger})$ where $\Lambda$ is the noise channel. The average fidelity $F_\text{avg}$, over all possible $n$-qubit states, given $\Lambda$ corresponding to the noisy application of $U$ is shown in \cite{MW23} to be
%\small
\begin{equation}
%\begin{strip}
%\begin{aligned}
\label{eqappRB1}
F_\text{avg}(U) =   \frac{1}{2^n(2^n+1)} 
 \sum_{\boldsymbol{i},\boldsymbol{j},\boldsymbol{i}^\prime\!,\boldsymbol{j}^\prime}\big ( \alpha^{U^{\dagger}}_{\ket{\boldsymbol{i}^\prime}\bra{\boldsymbol{j}^\prime}; \ketbra{\boldsymbol{i}}{\boldsymbol{j}}} +
\alpha^{U^{\dagger}}_{\ketbra{\boldsymbol{i}^\prime}{\boldsymbol{j}^\prime}; \ketbra{\boldsymbol{j}}{\boldsymbol{i}}} \big) \mathsf{Trace}\!\left[\ketbra{\boldsymbol{i}}{\boldsymbol{j}} \Lambda \circ \mathbf{U }(\ketbra{\boldsymbol{i}^\prime}{\boldsymbol{j}^\prime})\right] ,
%\end{aligned}
%\end{strip}
\end{equation}
%\normalsize
where

the coefficients $\alpha^{U^{\dagger}}_{\ketbra{\boldsymbol{i}^\prime}{\boldsymbol{j}^\prime}; \ketbra{\boldsymbol{i}}{\boldsymbol{j}}} := \bra{\boldsymbol{i}^\prime}U^\dagger\ket{\boldsymbol{i}}\bra{\boldsymbol{j}}U\ket{\boldsymbol{j^\prime}}\in \mathbb{C}$
%\alpha^{U^{\dagger}}_{\ketbra{\boldsymbol{i}^\prime}{\boldsymbol{j}^\prime}; \ketbra{\boldsymbol{i}}{\boldsymbol{j}}} \in \mathbb{C}$
depend only on the unitary $U$. This expression generally includes $2^{4n}$ terms, corresponding to $2^{4n}$ measurements. In the worst case, evaluating $F_\text{avg}(U)$ as above requires as many measurements as a full process tomography. However, for most gates of interest, a significant number of $\alpha$ coefficients will vanish, hence requiring many fewer measurements to evaluate the sum in Eq. (\ref{eqappRB1}). A more formal proof of Eq.~(\ref{eqappRB1}) will appear in \cite{MW23}, but we will now outline the key technical steps needed to arrive at Eq.~(\ref{eqappRB1}). 

We start by expanding $U|\psi\rangle=\sum_{\boldsymbol{i}}\beta_{\boldsymbol{i}} |\boldsymbol{i}\rangle$ in the basis $\{|\boldsymbol{i}\rangle\}$ for any initial state $\ket{\psi}$, with $\beta_{\boldsymbol{i}} \in \mathbb{C}$. Then, by plugging this into the expression of the final state fidelity $F_{\psi}(U)$ given in the main text, we obtain  %$F_{\psi}(U)=\sum_{\boldsymbol{i},\boldsymbol{i'}}\beta^*_{\boldsymbol{i'}}\beta_{\boldsymbol{i}}|\boldsymbol{i}\rangle \langle \boldsymbol{i^'} |$. 
\begin{equation*}
    F_{\psi}(U)=\sum_{\boldsymbol{i},\boldsymbol{i'},\boldsymbol{j},\boldsymbol{j'}}\beta^*_{\boldsymbol{i}}\beta^*_{\boldsymbol{i'}}\beta_{\boldsymbol{j}}\beta_{\boldsymbol{j'}}\langle \boldsymbol{i}| \Lambda(|\boldsymbol{j'}\rangle \langle \boldsymbol{i'}|)|\boldsymbol{j}\rangle,
\end{equation*}
and therefore
\begin{equation*}
    F_\text{avg}(U)=\sum_{\boldsymbol{i},\boldsymbol{i'},\boldsymbol{j},\boldsymbol{j'}}\mathbb{E}(\beta^*_{\boldsymbol{i}}\beta^*_{\boldsymbol{i'}}\beta_{\boldsymbol{j}}\beta_{\boldsymbol{j'}})\langle \boldsymbol{i}| \Lambda(|\boldsymbol{j'}\rangle \langle \boldsymbol{i'}|)|\boldsymbol{j}\rangle,
\end{equation*}
where the expectation values of the product of $\beta$ coefficients is
\begin{equation*}   
    \mathbb{E}(\beta^*_{\boldsymbol{i}}\beta^*_{\boldsymbol{i'}}\beta_{\boldsymbol{j}}\beta_{\boldsymbol{j'}}):=\int d\psi \beta^*_{\boldsymbol{i}}\beta^*_{\boldsymbol{i'}}\beta_{\boldsymbol{j}}\beta_{\boldsymbol{j'}}( \psi).
\end{equation*}
The quantities $\mathbb{E}(\beta^*_{\boldsymbol{i}}\beta^*_{\boldsymbol{i'}}\beta_{\boldsymbol{j}}\beta_{\boldsymbol{j'}})$ are the same as those in the expansion in the operator basis $\{|\boldsymbol{j} \rangle |\boldsymbol{j}^{'} \rangle \langle \boldsymbol{i}^{'} | \langle \boldsymbol{i}| \} $ of
\begin{equation*}
    \Pi:=\int d\psi (|\psi\rangle \langle \psi|)^{ \otimes 2}.
\end{equation*}
In addition, $\Pi$ is proportional to the projector onto the symmetric subspace of $(\mathbb{C}^{d})^{\otimes 2}$ with $d=2^n$ \cite{renes2004symmetric}, and this projector has a known expansion in the basis  $\{|\boldsymbol{j} \rangle |\boldsymbol{j}^{'} \rangle \langle \boldsymbol{i}^{'} | \langle \boldsymbol{i}| \} $ \cite{harrow2013church}. These key insights allow us to compute the coefficients 
$\mathbb{E}(\beta^*_{\boldsymbol{i}}\beta^*_{\boldsymbol{i'}}\beta_{\boldsymbol{j}}\beta_{\boldsymbol{j'}})$, to obtain 

\begin{equation*}
    F_\text{avg}(U)=\frac{1}{2^n(2^n+1)}\sum_{\boldsymbol{i},\boldsymbol{j}} \big(\langle \boldsymbol{i} | \Lambda (|\boldsymbol{j} \rangle \langle \boldsymbol{i}|) |\boldsymbol{j} \rangle+\langle \boldsymbol{i} | \Lambda (|\boldsymbol{i} \rangle \langle \boldsymbol{j}|) |\boldsymbol{j} \rangle\big).
\end{equation*}
The last steps needed to arrive at Eq. (\ref{eqappRB1}) are to note that $\langle \boldsymbol{i} | \Lambda (|\boldsymbol{j} \rangle \langle \boldsymbol{i}|)|\boldsymbol{j} \rangle=\textsf{Trace}(|\boldsymbol{j} \rangle \langle \boldsymbol{i}|\Lambda (|\boldsymbol{j} \rangle \langle \boldsymbol{i}|))$, and then to rewrite $\Lambda=\Lambda \circ \mathbf{U} \circ \mathbf{U}^{\dagger}$, and finally expand $\mathbf{U}^{\dagger}(|\boldsymbol{j}\rangle \langle \boldsymbol{i} |)$ in the basis $\{|\boldsymbol{i}^{'}\rangle \langle \boldsymbol{j}^{'} |\}$ to identify the $\alpha$ coefficients of Eq. (\ref{eqappRB1}).

Following this method, we can derive exact expressions of $F_\text{avg}$ for any gate unitary as a discrete sum of a finite number of terms. As outlined in the following sections, the exact form of $F_\text{avg}$ and the number of non-zero terms will depend on $U$, and these terms can be evaluated using a set of state preparation and measurement settings.

\bigskip

We note that there is in the literature another technique to evaluate $F_\text{avg}$ using fewer measurements than full process tomography. The approach experimentally implemented in \cite{o2004quantum} obtains $F_\text{avg}$ by first measuring an intermediate quantity called the entanglement fidelity \cite{nielsen2002simple,horodecki1999general,flammia2011direct}. In contrast, the symmetry-based method outlined here and used in the main text bypasses computing this intermediate quantity and directly evaluates $F_\text{avg}$ as a weighted summation of measurements. As discussed below, we notably find that the symmetry-based benchmarking approach requires half as many measurements to benchmark a CNOT gate as were required in \cite{o2004quantum}.

\bigskip 
\subsection{Average fidelity of a $T$-gate}
As an illustration of the method in action, we explicitly compute $F_\text{avg}(U)$ for the case of the $T$-gate,  which is a very important gate in the context of magic state distillation protocols for fault-tolerant quantum computing \cite{bravyi2005universal}. This gate is defined by the unitary transformation 
\begin{equation}
    \label{eqtgate}
    T =\begin{pmatrix}
    1 & 0 \\ 0 & e^{i \frac{\pi}{4}}
    \end{pmatrix}.
\end{equation}
In this case, the following relations can directly be verified by taking $n=1$ and replacing $U$ by $T$ to compute the $\alpha$ coefficients:
\begin{equation}
\begin{aligned}
    \label{eqappRB4}
    \alpha^{T^{\dagger}}_{\ketbra{0}{0};\ketbra{0}{0}}&=\alpha^{T^{\dagger}}_{\ketbra{1}{1};\ketbra{1}{1}}=1, \\
    \alpha^{T^{\dagger}}_{\ketbra{0}{1};\ketbra{0}{1}}&=e^{i \frac{\pi}{4}}, \,
    \alpha^{T^{\dagger}}_{\ketbra{1}{0};\ketbra{1}{0}}=e^{-\frac{i\pi}{4}}, \\
    \alpha^{T^{\dagger}}_{\ketbra{i}{j};\ketbra{i^\prime}{j^\prime}}&=0
    \, \, \mathrm{when} \, \,  i \neq i^\prime \, \mathrm{or} \,  j \neq j^\prime.
\end{aligned}
\end{equation}
Plugging these values into Eq.~(\ref{eqappRB1}) for $n=1$ gives
\begin{equation}
\begin{aligned}
    \label{eqappRB5}
    F_\text{avg}(U)&=\frac{1}{3}\mathsf{Trace}\left[\ket{0}\bra{0} \Lambda \circ \mathbf{T} (\ketbra{0}{0})\right] + \frac{1}{3}\mathsf{Trace}\left[\ketbra{1}{1} \Lambda \circ \mathbf{T} (\ketbra{1}{1})\right]\\
    &+ \frac{e^{i\frac{\pi}{4}}}{6}\left(\mathsf{Trace}\left[\ketbra{0}{1} \Lambda \circ \mathbf{T} (\ketbra{0}{1})\right]+ \mathsf{Trace}\left[\ketbra{0}{1} \Lambda \circ \mathbf{T} (\ketbra{1}{0})\right]\right)\\ \, &+ \frac{e^{-i\frac{\pi}{4}}}{6}\left(\mathsf{Trace}\left[\ketbra{1}{0} \Lambda \circ \mathbf{T} (\ketbra{0}{1})\right]+ \mathsf{Trace}\left[\ketbra{1}{0} \Lambda \circ \mathbf{T} (\ketbra{1}{0})\right]\right).
\end{aligned}
\end{equation}
Then, we can note that 
\begin{equation}
\begin{aligned}
    \label{eqappRB6}
    \ketbra{0}{1}&=\ketbra{+}{+} + i\ketbra{+_i}{+_i}-\frac{1+i}{2}I \\
    \ketbra{1}{0}&=\ketbra{+}{+} - i\ketbra{+_i}{+_i}-\frac{1-i}{2}I,
\end{aligned}
\end{equation}
where $\ket{+}=(\ket{0}+\ket{1})/\sqrt{2}$, $\ket{+_i}=(\ket{0}+i\ket{1})/\sqrt{2}$, and $I=\ketbra{0}{0}+\ketbra{1}{1}$. In addition, since $\Lambda \circ \mathbf{T}$ is a CPTP map, then $ \Lambda \circ \mathbf{T} (I)=I$, and consequently $\mathsf{Trace}\left[\ketbra{1}{0}\Lambda \circ \mathbf{T} (I)\right]=\mathsf{Trace}\left[\ketbra{0}{1}\Lambda \circ \mathbf{T} (I)\right]=0$. Thus, we can write the average fidelity in terms of four combinations of state preparations $\mathcal{I}=\{\ket{0},\ket{1},\ket{+},\ket{+_i}\}$ and measurements $\mathcal{M}=\{\ketbra{0}{0},\ketbra{1}{1},\ketbra{+}{+}\}$ that can be implemented on Ascella:
\begin{equation}
    \begin{aligned}
        F_\text{avg}(U)&=\frac{1}{3}\mathsf{Trace}\left[\ketbra{0}{0} \Lambda \circ \mathbf{T} (\ketbra{0}{0})\right] +\frac{1}{3}\mathsf{Trace}\left[\ketbra{1}{1} \Lambda \circ \mathbf{T} (\ketbra{1}{1})\right]\\ &- \frac{2}{3 \sqrt{2}}\mathsf{Trace}\left[\ketbra{+}{+}\Lambda \circ \mathbf{T} (\ketbra{+_i}{+_i})\right] \\  &+\frac{2}{3 \sqrt{2}}\mathsf{Trace}\left[\ketbra{+}{+} \Lambda \circ \mathbf{T} (\ketbra{+}{+})\right].
    \end{aligned}
\end{equation}

\subsection{Multi-qubit gates}
To evaluate Eq. (\ref{eqappRB1}) for any given gate unitary $U$, we must generalize the approach applied above. This can be done by first pre-computing $\Lambda \circ \mathbf{U}$ in terms of single-qubit Pauli operators $S_j\in\{I,X,Y,Z\}$, and then explicitly evaluating $F_{\mathrm{avg}}(U)$ to significantly reduce the required set of measurements that will ultimately be performed by Ascella. This approach provides a solution for $F_{\mathrm{avg}}(U)$, but it is not necessarily the optimal way to determine the required measurement settings.

To evaluate the process map in terms of measurements that can be performed by Ascella, we can choose to prepare each qubit $j$ independently following $\ket{\psi_j}\in\mathcal{I}$ as for the $T$-gate and measure each qubit independently following $S_j\in\{I,X,Y,Z\}$. This choice of state preparations and measurements also may not be optimal to minimize the number of measurements for any given gate, but it guarantees a system of $2^{4n}$ linearly independent equations describing an $n$-qubit noisy gate:
\begin{equation}
\label{paulicorrelations}
\braket{S}_{\Psi}=\mathsf{Trace}\left[S~\Lambda\circ \mathbf{U}(\ket{\Psi}\bra{\Psi})\right],
\end{equation}
where $S=S_1\otimes S_2\cdots S_n$ and $\ket{\Psi}=\ket{\psi_1}\otimes\ket{\psi_2}\dots\ket{\psi_n}$. 

For $n\leq 3$, this system can be solved directly using symbolic packages or software such as $\mathsf{Mathematica}$. Solving $4$-qubit gates or larger would require more advanced numerical methods or analytic simplification to circumvent memory limitations. Hence, although the approach to obtain $F_{\mathrm{avg}}(U)$ is general for any gate and results in equal or fewer required measurements compared to a full process tomography, pre-computing these required measurements is still a hard problem.

% \subsection{Measurement settings for 2- and 3-qubit gates}
Evaluating Eq.~(\ref{paulicorrelations}) for the CNOT and Toffoli gates provides all elements of the error process $\Lambda$ in terms of Pauli correlations. By evaluating $F_{\mathrm{avg}}(U)$ using the methods of \cite{MW23}, we obtain a set of 58 (593) correlations for the CNOT (Toffoli) gate. The set of correlations $m_i$ and their weight $w_i$ required for the evaluation of $F_{\mathrm{avg}}(U)$ for a CNOT gate are given in Table \ref{fidelityexpressionsCNOT}. These solutions represent a significant reduction over the number of correlations that would be necessary to perform full arbitrary process tomography, which is $2^{4n}=256$ ($4096$) for the CNOT (Toffoli) gate.

Since some of the correlations are measured among fewer than $n$ qubits, it is possible to further reduce the number of measurement settings by tracing out some qubits from the measurements obtained from higher-order correlations. For example, $\braket{IX}$ can be evaluated from the same data used to compute $\braket{ZX}$ by tracing out the first qubit. In this case, assuming all $I$ measurements can be evaluated by recycling the $Z$ measurement data, the number of measurements is reduced to $36$ ($340$) for the CNOT (Toffoli), corresponding to $86\%$ ($92\%$) fewer measurements than for a full process tomography. Recycling the $X$ measurement further reduces this to $34$ settings for the CNOT gate. Notably, $34$ measurement settings are fewer than half of the $71$ settings previously used to benchmark a linear-optical CNOT gate in \cite{o2004quantum}.

\subsection{Benchmarking results}

We applied the symmetry-based benchmarking method to gates implemented by Ascella and also gates implemented on other online quantum computing platforms.

\subsubsection{Ascella}

For each individual state preparation and measurement (SPAM) configuration, the transpilation process converges to a high-fidelity implementation of the desired unitary. However, one of the assumptions needed to apply symmetry-based benchmarking is that the gate unitary remains unchanged for each SPAM configuration. As a result, re-transpiling the unitary for each setting can introduce a small systematic, but random, error in the estimate of the average gate fidelity.

When benchmarking the $T$-gate naively, we find that the re-transpilation error can occasionally cause the measured average gate fidelity to exceed $1$ by up to $0.1\%$, implying that the re-transpilation error is on the same order of magnitude as the $T$-gate error for Ascella. To remove this systematic error, we fix the voltages applied to the specific part of the chip implementing the $T$-gate so that transpilation process only optimizes the SPAM procedure. As a result, the measured average gate fidelity given in the main text is less than $1$ to within the measurement precision, but it also no longer fully benefits from the advantage of the machine-learned transpilation process.

We also observe that implementing multiple $T$-gates in a row, up to $4$ $T$-gates each implemented by a separate part of the chip, does not significantly decrease the measured average single-qubit gate fidelity. This suggests that the dominant contribution to the remaining $T$-gate error of $0.4\%$ is likely SPAM error.

Since the systematic error caused by re-transpilation is much smaller than the CNOT and the Toffoli gate errors and on the same order of magnitude as the measurement precision we apply the transpilation process to the entire chip for those cases. It is worth noting that implementing the Toffoli gate already saturates all $12$ modes of the chip, meaning that it is not possible to implement SPAM without re-transpiling the part of the chip implementing the Toffoli gate.

\subsubsection{Other online platforms}

\begin{figure}
\begin{tabular}{cc}
    \includegraphics[width=0.3\linewidth]{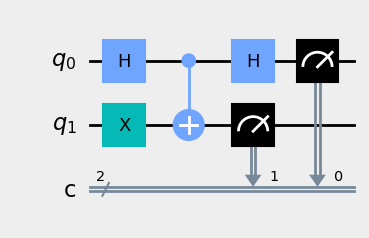} 
    & \includegraphics[width=0.37\linewidth]{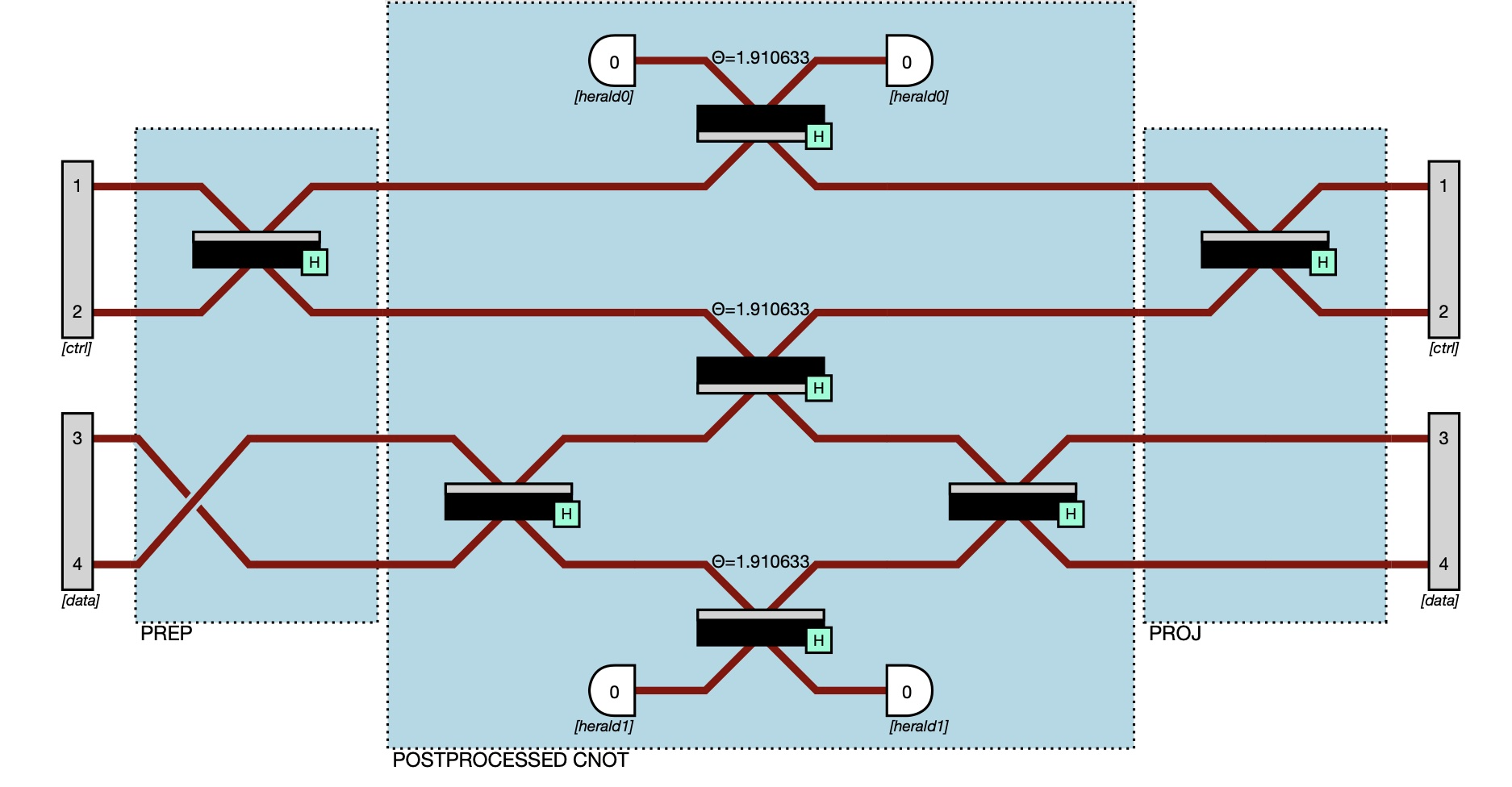} 
    
\end{tabular}
    
    \caption{One of the $36$ circuits ({\tt +1:XZ}), quantum circuit and linear optics circuit equivalent contributing to the $2$-qubit CNOT fidelity measurement}
    \label{fig:circuit_+1-XZ}
\end{figure}

To place Ascella in context, we also apply the same symmetry-based benchmarking for gates implemented by several other online quantum computing platforms (see Table \ref{tab:fgatecomparison}).  We show in Figure~\ref{fig:circuit_+1-XZ} an example of a circuit for running benchmarking on these platforms compared to the equivalent circuit on Ascella.

\begin{table}[]
    \centering
    \begin{tabular}{p{5cm}|c|c|p{5cm}}
    Platform (Device) & Gate & $F_\text{avg}$ (\%) & Date - Benchmark Details\\ \hline\hline
         Quandela (Ascella) & $T$-gate & $99.6\pm 0.1$ & {\scriptsize 2023/05/31 -- average and standard deviation on $5\times$ 1M-sample measurements, for 14 different gate locations on chip} \\
                 & CNOT & $93.8\pm 0.6$ & {\scriptsize 2023/03/20-2023/05/07 --- average and standard deviation of 114 consecutive 100k-sample measurements over 46 days} \\
                 & Toffoli & $86\pm 1.2$ & {\scriptsize 2023/01/06 -- calculated on 100000-sample tasks} \\
         \hline
         IonQ ({\tt AWS ionq.qpu})    & $T$-gate & $99.6 \pm 1$ & {\scriptsize 2022/12/16 -- calculated on 4096-sample tasks}\\
                 & CNOT & $91.7 \pm 1.5$  & {\scriptsize 2022/12/17 -- calculated on 4096-sample tasks}\\
                 & Toffoli  & $90\pm 3.1$ & {\scriptsize2023/01/18 -- calculated on 256-sample tasks} \\ \hline
         Rigetti ({\tt AWS rigetti.aspen-11}) & $T$-gate & $88.7 \pm 1$ & {\scriptsize 2022/12/16 -- calculated on 4096-sample tasks}\\
                 & CNOT & $71.2 \pm 1.5$ & {\scriptsize 2022/12/17 -- calculated on 4096-sample tasks}\\ \hline
         IBM ({\tt Quito} or {\tt Belem} depending on availability)     & $T$-gate & $96 \pm 1.5$ & {\scriptsize 2022/12/16 -- calculated on 4096-sample tasks} \\
                 & CNOT & $86.4 \pm 1.5$ & {\scriptsize 2022/12/17 -- calculated on 4096-sample tasks}
    \end{tabular}
    \caption{Average gate fidelity $F_\text{avg}$ obtained from symmetry-based benchmarking applied to devices on other online quantum computing platforms. Margin of uncertainty is based on standard deviation assuming Poisson counting statistics unless otherwise stated. Note that this benchmarking procedure is not robust to SPAM errors, which may be a significant factor for some devices. In addition, the values returned by other online platforms may be post-processed using error mitigation techniques or subject to systematic errors analogous to the re-transpilation errors on Ascella.}
    \label{tab:fgatecomparison}
\end{table}

\begin{table}[h]
\centering
    $\small
    \begin{array}{cl||cl||cl||cl||cl||cl}
w_i & m_i & w_i & m_i & w_i & m_i & w_i & m_i & w_i & m_i & w_i & m_i \\\hline\hline
 1 & \text{00:II} & 1 & \text{01:II} & 2 & \text{0+:IX} & -1 & \text{10:YY} & 1 &
   \text{11:YY} & 2 & \text{+0:YY} \\
 -1 & \text{00:IX} & -1 & \text{01:IX} & 2 & \text{0+:XI} & -1 & \text{10:ZI} & -1 &
   \text{11:ZI} & 2 & \text{+1:XI} \\
 1 & \text{00:IZ} & -1 & \text{01:IZ} & 2 & \text{0+:ZX} & 1 & \text{10:ZX} & 1 &
   \text{11:ZX} & -2 & \text{+1:XX} \\
 -1 & \text{00:XI} & -1 & \text{01:XI} & -2 & \text{0i:XZ} & 1 & \text{10:ZZ} & -1 &
   \text{11:ZZ} & -2 & \text{+1:YY} \\
 1 & \text{00:XX} & 1 & \text{01:XX} & 1 & \text{10:II} & 1 & \text{11:II} & 2 &
   \text{1+:IX} & -4 & \text{++:XI} \\
 1 & \text{00:XZ} & 1 & \text{01:XZ} & -1 & \text{10:IX} & -1 & \text{11:IX} & 2 &
   \text{1+:XI} & -2 & \text{i0:XZ} \\
 -1 & \text{00:YY} & 1 & \text{01:YY} & -1 & \text{10:IZ} & 1 & \text{11:IZ} & -2 &
   \text{1+:ZX} & -2 & \text{i1:XZ} \\
 1 & \text{00:ZI} & 1 & \text{01:ZI} & -1 & \text{10:XI} & -1 & \text{11:XI} & -2 &
   \text{1i:XZ} & 4 & \text{ii:XZ} \\
 -1 & \text{00:ZX} & -1 & \text{01:ZX} & 1 & \text{10:XX} & 1 & \text{11:XX} & 2 &
   \text{+0:XI} & \text{} & \text{} \\
 1 & \text{00:ZZ} & -1 & \text{01:ZZ} & 1 & \text{10:XZ} & 1 & \text{11:XZ} & -2 &
   \text{+0:XX} & \text{} & \text{} \\
\end{array}
$
    \caption{List of $58$ weights $w_i$ and correlations $m_i$ used to evaluate the average fidelity $F_\text{avg}(U)=(1/40)\sum_iw_im_i$ of a CNOT gate. Correlations are labelled by $ab$:$xy$ where $a,b\in\{0,1,+,\text{i}\}$ represent state preparation ($\{\ket{0},\ket{1},\ket{+},\ket{+_i}\}$ respectively), and $x,y\in\{\text{I,X,Y,Z}\}$ represent measurements ($\{I,X,Y,Z\}$ respectively).}
    \label{fidelityexpressionsCNOT}
\end{table}

\section{Boson Sampling} \label{app:boson_sampling}

In this section we study boson-sampling with photon loss. With a $6$-photon input state, we acquire respectively $295 \cdot 10^{9}$, $41.5 \cdot 10^{9}$, $3.07 \cdot 10^{9}$, $110 \cdot 10^{6}$, and $1.78 \cdot 10^{6}$ samples for $1$-, $2$-, $3$-, $4$-, and $5$-photon coincidences. We compute the total variation distance $D=\frac{1}{2}\sum_i|p_i-q_i|$  where $\{p_i\}$ and $\{q_i\}$ are the ideal and experimental output probability distributions respectively for output states with $2$, $3$, $4$ and $5$ photons. The results are reported in Tab.~\ref{tab:boson_sampling}. For $>2$ photon loss the experimental output probability is dominated by loss, and no longer describe the ideal distribution.

\begin{table}[h]
    \centering
    \begin{tabular}{c|c|c}
        $N$-photon loss & Fidelity & Distance \\
        \hline
        0- & 0.97±0.03 & 0.16±0.02 \\
        \hline       
        1- & 0.989±0.002 & 0.118 \\
        \hline
        2- & 0.9950±0.0002 & 0.143 \\
        \hline
        3- & 0.99625±3e-05 & 0.225\ \\
        \hline
        4- & 0.997333±6e-06 & 0.40 \\
        \hline
    \end{tabular}
    \caption{Fidelity and total variation distance between the experimental data and the ideal output probability for all $N$-photon outcomes.} \label{tab:boson_sampling}
\end{table}

To validate that our boson-sampler device is functioning correctly, we use a phenomenological approach that models all sources of noise in the experimental apparatus and demonstrate that we reach a remarkable overlap between all threshold statistics and our simulations. Using \textit{Perceval}, we develop a realistic simulator for our boson-sampler. We use the phenomenological model first introduced in~\cite{pont2022} of our single-photon source to account for the partial distinguishability of the multiphoton state, the imperfect single-photon purity and the optical losses. Then, we account for the error related to the transpilation of the unitary matrix and the imperfect implementation of the physical phases with the thermo-optic phase shifters. The simulations gives a total variation distance of $D_{\text{simu}}=0.132$ and a fidelity of $F_{\text{simu}}=0.978$. The simulations are compatible with the experimental data, which shows that our realistic simulator truthfully describes the boson-sampling device.

\section{Classification}\label{app:classification}

Quantum algorithms for classification on near-term devices have been explored through a variety of approaches \cite{Li_2021}, although most results are supported by numerical simulations and not implemented in the lab. This is the case for instance of Ref.~\cite{Gan_2022}, where the authors present an ansatz for a variational quantum algorithm that is native to photonics. They study the expressivity of the resulting model through theory and numeric simulations. The circuit is made of two trainable blocks with a data encoding block in the middle. This data encoding block consists of phase shifters. For a $k$-dimensional data point $\Vec{x} = (x_i, \dots, x_k)$, each feature $x_i$ is encoded into the phase of a phase shifter. The two trainable blocks are beamsplitter meshes that implement unitary operations, for instance through the encoding of \cite{Reck_1994} or \cite{Clements_2016}. The model is studied within the Fock space, considering input states $\ket{n^{in}_{1}, \dots, n^{in}_{m}}$ and output states $\ket{n^{out}_{1}, \dots, n^{out}_{m}}$. The total number of photons is denoted $n$ and the number of modes is denoted $m$. 

The authors of Ref.~\cite{Gan_2022} show that the output of the circuit, i.e. the model, can be expressed as $f_{\boldsymbol{\theta}}(x) = \sum_{\omega} c_\omega(\boldsymbol{\theta}, \boldsymbol{\lambda}) e^{ix\omega}$,
where the frequencies $\omega$ depend on the number of photons $n$ input in the circuit, and the Fourier coefficients $c_\omega(\boldsymbol{\theta}, \boldsymbol{\lambda})$ depend on the measurement parameters $\boldsymbol{\lambda}$ and the chip parameters $\boldsymbol{\theta}$. 

We can see how interesting it is to use a photonic encoding and in particular to exploit the Fock space from the perspective of expressivity: by adding more photons, more terms will be added to the Fourier series and the resulting model will be more expressive, without increasing the complexity of the circuit or the number of modes. However, it is important to note that this is only possible if PNR detectors are available, so that the output state $\ket{n^{out}_{1}, \dots, n^{out}_{m}}$ can be resolved beyond $n^{out}_{i} = 1$.

\begin{figure}[h]
    \centering
    \includegraphics[width=0.5\textwidth]{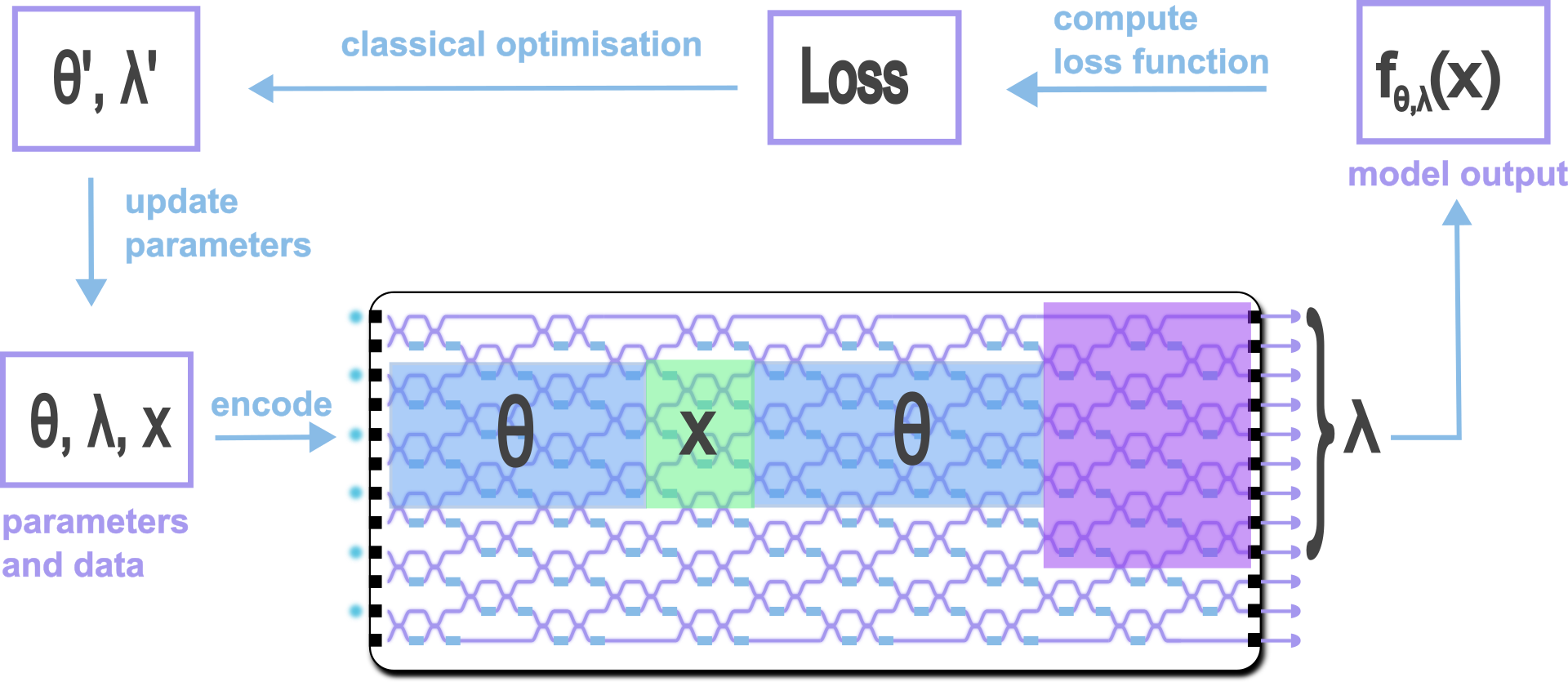}
    \caption{Representation of our variational quantum classifier on Ascella for the classification algorithm. The trainable blocks parameterized by $\boldsymbol{\theta}$ are depicted in blue. The data encoding block is in green. The purple block represents the pseudo-PNR layer, which incorporates four extra modes into the circuit. The model is computed by assigning $\boldsymbol{\lambda}$ parameters to the outputs at the detectors. All parameters are updated via classical optimization.}
    \label{fig:classifier_ansatz}
\end{figure}

Taking inspiration from the results of \cite{Gan_2022}, we design an ansatz for the variational quantum classifier on Ascella containing two parameterizable blocks with a data-encoding block in between. We select modes $3$ to $7$ on Ascella and construct the first parameterized block using $16$ of the reconfigurable thermo-optic phase shifters. The data-encoding block follows, where $4$ phase shifters acting on modes $4$ to $7$ encode the $4$ features of each IRIS data point. We then implement another parameterized block with $16$ phase shifters. The remaining phases on the chip are either set to $0$, when we do not wish to add any extra trainable or encoding phases, or to $\pi$, when an interferometer needs to be fully reflective so that no photon escapes to the other modes of the chip. Our ansatz is shown in Fig.~\ref{fig:classifier_ansatz}.

We input into the circuit the $3$-photon Fock state $\ket{\psi_{in}} = \ket{0010101000000}$ defined over the $12$ modes of Ascella: one photon enters modes $3$, $5$ and $7$ of the chip. We observe different output states $\ket{\psi_{out}}$ depending on the photon counts observed in the detectors. The resulting model takes the form $f_{\boldsymbol{\theta}, \boldsymbol{\lambda}}(\boldsymbol{x}) = \bra{\psi_{out}} \mathcal{U}^{\dagger}(\boldsymbol{x}, \boldsymbol{\theta}) \mathcal{M}(\boldsymbol{\lambda}) \mathcal{U}(\boldsymbol{x}, \boldsymbol{\theta}) \ket{\psi_{in}}.$ The weights $\boldsymbol{\lambda}$ are assigned to each possible output state observed at the detection step and thus define the observable $\mathcal{M}$ that we are effectively measuring. The operator $\mathcal{U} (\boldsymbol{x}, \boldsymbol{\theta})$ describes the parameterized and data-encoding blocks. Following the variational approach, we train the model by optimizing classically over the phases $\boldsymbol{\theta}$ from the parameterized blocks, as well as over the weights $\boldsymbol{\lambda}$.

We demonstrate pseudo photon-number resolution (PNR) partially, on the first and on the last detectors. To this end, we set four phase shifters to $\pi/2$ in the final layer of the chip in order to redirect a portion of the photons from modes $3$ and $7$ into modes $1$, $2$ and $8$, $9$ respectively. We can then reinterpret the detection counts: for instance, observing photons in modes $1$, $2$ and $3$ in our scheme corresponds to observing a three-photon count in mode $3$ if we had PNR detectors. Note that implementing this partial pseudo-PNR adds expressivity to our model, as it increases the space of possible outcome states and thus the amount of $\boldsymbol{\lambda}$ parameters over which we optimize.

For the optimization, we found that performing a see-saw was the most efficient option. This means separating the chip parameters $\boldsymbol{\theta}$ from the observable parameters $\boldsymbol{\lambda}$ into two loops, and finding the best $\boldsymbol{\lambda}$ for each set of values of $\boldsymbol{\theta}$. The efficiency comes from the fact that tuning one or other set of parameters is not equally costly: changing the $\boldsymbol{\theta}$ requires re-configuring the chip, while modifying the $\boldsymbol{\lambda}$ only involves classical post-processing. We used Gaussian processes to optimize the $\boldsymbol{\theta}$ and for each iteration we optimized over the $\boldsymbol{\lambda}$ using the Nelder-Mead method.

A note about optimizing over the $\boldsymbol{\lambda}$ parameters: if we were to choose a fixed observable for the variational algorithm and dismiss the $\boldsymbol{\lambda}$ parameters completely, we may not obtain the same performance for the classifier. Indeed, we would not only have fewer parameters for the optimization but the remaining parameters would be the ones most sensitive to the noise of the experiment. Nevertheless, it is reasonable to grant degrees of freedom to the choice of the observable, so we choose to optimize over the $\boldsymbol{\lambda}$ as in \cite{Gan_2022}.

\begin{figure}[h]
\begin{tabular}{cc}
    \begin{minipage}{0.5\columnwidth}
            \centering
          \includegraphics[width=\textwidth]{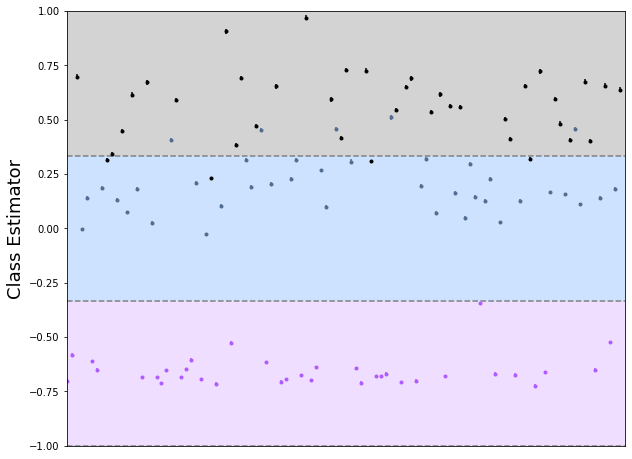}
         \caption{Training dataset}
        %\label{subfig:g2}
\end{minipage}
            &       
      \begin{minipage}{0.5\columnwidth}
             \centering
        \includegraphics[width=\textwidth]{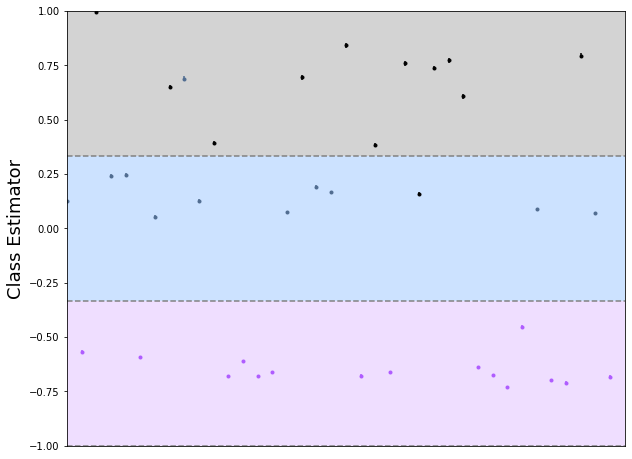}
         \caption{Test dataset}
        %\label{subfig:hom}
\end{minipage}      
 \end{tabular}
   \caption{Value of the classification estimator, i.e.\ the model output, displayed for each data point in the train and the test set. The color of the data point indicates its true label while the background color indicates the predicted label. The error bars are obtained via Poissonian statistics from the number of samples used in each run to compute the value of the estimator. The points are simply ordered in the same way as in the dataset.}
        \label{fig:classifier_with_error}    \end{figure}

% Results including error bars
In the main text, we summarized the performance of the model using confusion matrices. Fig.~\ref{fig:classifier_with_error} shows an alternative display of the results, where the classification estimator is included for each prediction in the dataset along with an error. We evaluated this error knowing that we used $50000$ samples for each run of the experiment. Note also that we adapted the definition of the classification estimator of \cite{Gan_2022} to our case of multi-class classification.

\section{VQE}
\label{app:VQE}

\begin{figure}
    \centering
    \includegraphics[width=0.5\textwidth]{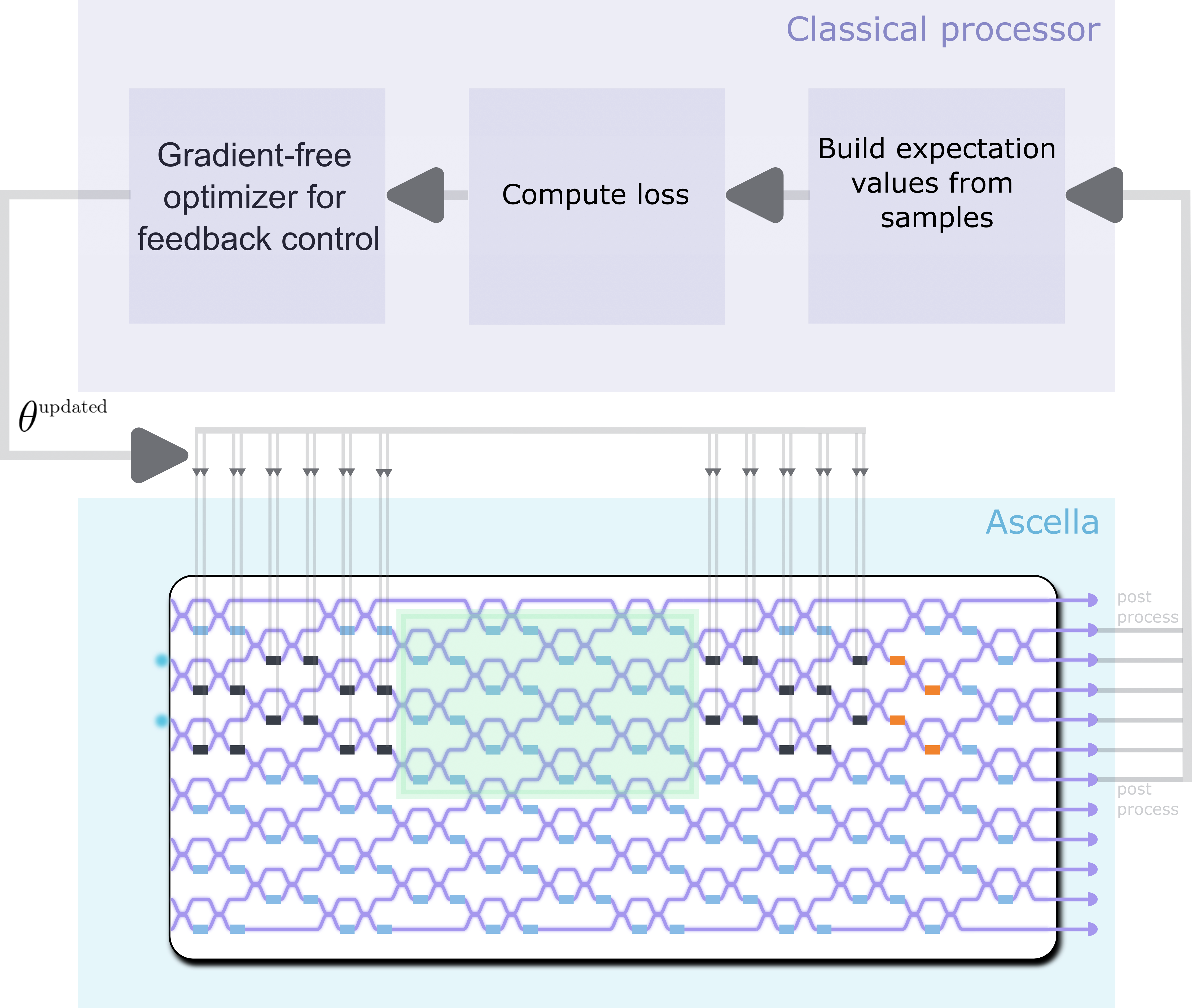}
    \caption{Procedure for the hybrid photonic VQE algorithm we implement on Ascella. It comprises a QPU block on the bottom that enables the energy of the chosen system to be estimated. The ansatz is parameterized via thermo-electric phase shifters highlighted in black. The green block is a postselected Ralph CNOT \cite{ralph2002linear}. The phase shifter in orange allows selection of the measurement bases required to reconstruct the correct Pauli terms from the qubit Hamiltonian \eqref{eqapp:qubithamiltonian}. The outputs from Ascella are fed into a classical block (at the top) which reconstructs the energy. Then it implements a feedback loop back into Ascella via a gradient-free optimizer in order to optimize the angles of the phase shifters to obtain an ansatz closer to the ground state.}
    \label{fig:procedure_VQE}
\end{figure}

\begin{figure}[h!]
\centering
         \centering
         \includegraphics[width=0.5\textwidth]{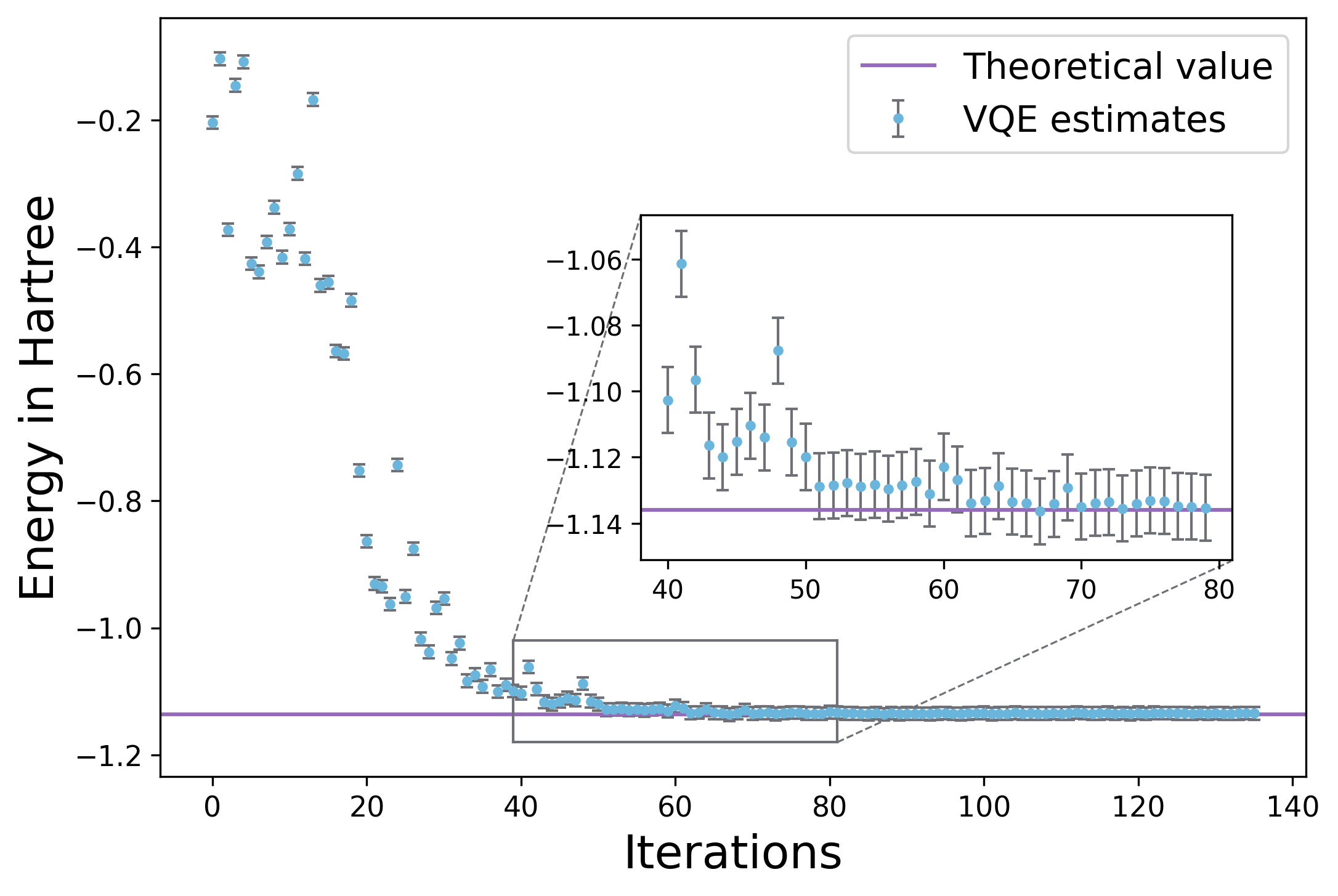}
         \caption{Evolution of the ground state energy of H$_2$ obtained by the VQE procedure on Ascella with the number of iterations.}
         \label{fig:evolution_VQEenergy}
\end{figure}

The Variational Quantum Eigensolver (VQE) \cite{peruzzo2014variational} for finding ground state energies for a target Hamiltonian can be broken into several steps as follows:
\begin{itemize}
    \item[1.] Find a Hamiltonian formulation suitable for the problem at hand. To do so, we use the symmetry-conserving Bravyi-Kitaev transformation \cite{bravyi2017tapering} (available through the \textit{OpenFermion} \cite{mcclean2020openfermion} python package). This provides two very useful pieces of information:
    \begin{itemize}
        \item The number of qubits necessary to run the VQE.
        \item A description of the Hamiltonian in terms of Pauli words which then define coefficients to associate to each measurement basis. We can then construct the expectation value based on this description. 
    \end{itemize}
    \item[2.] Prepare the ansatz. We need a parametrizable circuit expressive enough to be able to reach (a good approximation of) the ground state. 
    \item[3.] Rotate the ansatz into the basis for each Pauli word operator present in the molecular Hamiltonian, then take a number of samples. Restore the ideal sample counts through error mitigation.
    \item[4.] Optimize the parameters via a classical procedure (COBYLA algorithm) based on the expectation value computed from the error-mitigated output of the QPU. A classical feedforward loop updates the parameters of the ansatz to reach a lower energy. The evolution of the ground state energy of H$_2$ with respect to the iterations is represented in Fig.~\ref{fig:evolution_VQEenergy}.
\end{itemize}

\subsection{Hamiltonian description}

In order to calculate expectation values of our H$_2$ Hamiltonian we need to transform the second-quantized version of the Hamiltonian into a qubit basis. This can be done with a number of different transformations such as Jordan-Wigner, but here we use the Symmetry-Conserving Bravyi-Kitaev transform as described in \cite{bravyi2017tapering,seeley2012bravyi}. This involves first reordering the electronic orbitals and then using the Bravyi-Kitaev binary tree mapping.
%(the set of indices of the qubits that must be updated when the occupation of some orbital $j$ is modified), 
%%the parity set $P(j)$,
%(the set of indices of the qubits needed to determine the parity of orbitals with index lower than $j$) 
%%and the remainder set $R(j)$.
%Above, $U(j)$ denotes the update set --- the set of qubits that must be updated when the occupation of some orbital $j$ is changed. The parity set, $P(j)$, is the set of qubits needed to determine the parity of the set of orbitals with index lower than $i$. Finally, the remainder set $R(j)$ is defined by: $R(j)=P(j) \backslash F(j)$ ; the elements of $P(j)$ that are not in $F(j)$ (see \cite{bravyi2017tapering,seeley2012bravyi}).
This gives a Hamiltonian which consists of four states that can be assigned in the following way: qubit $1$ corresponds to spin-up on the first site, qubit $2$ to spin-up on the second site, qubit $3$ to spin-down on the first site, and qubit $4$ to spin-down on the second site. 
% \begin{equation}
% \begin{aligned}
%  \hat{\mathcal{H}}_{H_2} =& \,  \alpha_0 \mathbb{I}+\alpha_1 X_0 Z_1 X_2 + \alpha_3 X_0 Z_1 X_2 Z_3 + \alpha_4 X_0 X_2 + \alpha_5 X_0 X_2 Z_3 \\
%  +& \, \alpha_6 Z_0 + \alpha_7 Z_0 Z_1 + \alpha_8 Z_0 Z_1 Z_2 + \alpha_9 Z_0 Z_1 Z_2 Z_3 + \alpha_{10} Z_0 Z_2 \\
%  +& \, \alpha_{11} Z_0 Z_2 Z_3 + \alpha_{12} Z_1 + \alpha_{13} Z_1 Z_2 Z_3 + \alpha_{14} Z_1 Z_3 + \alpha_{15} Z_2
% \end{aligned}
% \end{equation}
% The constants $\alpha_{i}$ are calculated from the one and two electron integrals of the system $h_{pq}$ and $h_{pqrs}$ respectively and vary with different values of bond length of the molecule. 
The Hamiltonian can be further reduced with the symmetries first derived in \cite{o2016scalable}.
There it was noted that the Hamiltonian acts off-diagonally on only two qubits, those indexed $0$ and $2$.
The simulation is begun in the Hartree-Fock state
which stabilizes qubits $1$ and $3$ so that they are never flipped
throughout the simulation. This symmetry can be used to reduce the Hamiltonian of interest to the
following effective Hamiltonian which acts only on
two qubits:

\begin{equation}
\hat{\mathcal{H}}_{\mathrm{qubit}}= \alpha \mathbb{I}\mathbb{I} +\beta Z \mathbb{I}+\gamma \mathbb{I} Z+\delta Z Z +\mu X X
\label{eqapp:qubithamiltonian}
\end{equation}
% \begin{equation}
% \hat{\mathcal{H}}_{\mathrm{qubit}}= \alpha \mathbb{I}_0 \otimes \mathbb{I}_1+\beta Z_0 \otimes \mathbb{I}_1+\gamma \mathbb{I}_0 \otimes Z_1+\delta Z_0 \otimes Z_1 +\mu X_0 \otimes X_1
% \label{eqapp:qubithamiltonian}
% \end{equation}
The constants vary with the choice of bond length and can be found in Table~\ref{tab:Ham_table}. 
The procedure described above is implemented in \textit{OpenFermion} \cite{mcclean2020openfermion}.  Thus finding the ground state energy for H$_2$ for varying bound length can be performed
\begin{itemize}
    \item with $2$ qubits (providing we have an expressive enough ansatz -- this is what we tackle in the next subsection),
    \item by building the expectation values for the measurements $ZZ$ and $XX$. This is done in a single job on the QPU by tuning the phase shifter in orange in Fig.~\ref{fig:procedure_VQE} with micro-increments which allows only a few phases on the Ascella chip to be changed and avoids losing time with full reconfiguration of the chip. 
\end{itemize}

\subsection{Ansatz preparation}

\begin{figure}[h!]
    \centering
    \scalebox{.9}{
        \begin{quantikz}
            \lstick{$\ket{0}$} & \gate{R_Y(\theta_1)} & \ctrl{1} & \gate{R_X(\theta_2)} & \gate{R_Z(\theta_4)} & \gate{R_X(\theta_6)} & \qw \\
            \lstick{$\ket{0}$} & \qw & \targ{} & \gate{R_X(\theta_3)} & \gate{R_Z(\theta_5)} & \gate{R_X(\theta_7)} & \qw
        \end{quantikz}}
    \caption{Gate-based version of an ansatz circuit with $7$ parameters controlling $7$ parameterized rotations capable of generating any $2$-qubit state.}
    \label{fig:gatebasedVQE}
\end{figure}
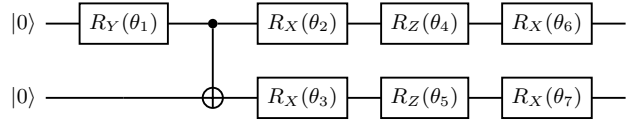

In the present context, the idea behind using the VQE algorithm is to produce a parameterizable ansatz expressive enough to get very close to a $2$-qubit ground state of the desired Hamiltonian in order to obtain the ground state energy. The gate-based circuit below (see Fig.~\ref{fig:gatebasedVQE}) can generate any $2$-qubit state by Schmidt decomposition (one $R_X$ rotation at the end can be removed in principle since it amounts to removing the global phase).  

This wave ansatz can be implemented photonically. To deal with noise and because using more parameters can be helpful for converging faster to the ground state energy, we use the ansatz represented in Fig.~\ref{fig:procedure_VQE} where we path encode the $2$ qubits with one photon per pair of modes $2-3$ and $4-5$ (the first mode being the $0^\text{th}$ one). This comprises $20$ tunable phase shifters. Modes $1$ and $6$ are used as ancillary mode for the postselected Ralph CNOT. The CNOT is constructed from our transpilation algorithm. 
We use parametrizable thermo-electric phase shifter (represented in black on Fig.~\ref{fig:procedure_VQE}) to control the optical index in the waveguide and thus tune the phase of the photons so that we can achieve any $2$-qubit state.

\subsection{Error Mitigation}

We use the quantum error mitigation (QEM)  technique proposed in \cite{ur2021entanglement} and first experimentally demonstrated in \cite{lee2022error} to more consistently converge to the correct ground state energy by mitigating state preparation and measurement (SPAM) errors.
The idea is to mitigate the errors arising
from noisy evolution due to thermal noise from the heated phase shifters as well as from the error caused by changing the measurement basis.
As detailed in the Methods section, we want to compute left-stochastic matrices $\Gamma_b$ (for each measurement basis $b$) such that
\begin{equation} \label{eqapp:errormitvqe}
    q = \Gamma_b p,
\end{equation}
with  $q$ the noisy output probability and $p$ the ideal noiseless output probability. In our case, we have two such $\Gamma_b$, which corresponds to a measurement basis in Eq.~\eqref{eqapp:qubithamiltonian}: $XX$ and $ZZ$ (as $\mathbb I \mathbb I$, $\mathbb I Z$ and $Z \mathbb I$ can be obtained from $ZZ$ by classical post-processing).

We construct $\Gamma_{b}$ experimentally as follows: 
$(\Gamma_b)_{ij} = \vert \bra \psi_i^b b \ket \psi_j^b \vert^2$ is the probability of obtaining the $i$$^\text{th}$ eigenvector of $b$ $\ket \psi_i^b$ when the $j$$^\text{th}$ eigenvector $\ket \psi^b_j$ is prepared and measurement observable $b$ is performed. For low enough SPAM errors, each $\Gamma_b$ is a diagonally dominant matrix and we can retrieve the idealized probability distribution by inverting $\Gamma_b$ in Eq.~\eqref{eqapp:errormitvqe}.
From this procedure we get the two desired matrices as:
\begin{equation}
{\footnotesize
\begin{aligned}\label{eq:Gammas} \centering
\Gamma_{ZZ} & =
\begin{bmatrix}
9.99999952e-01 & 3.09568451e-02 & 3.09568451e-02& 1.54929555e-09\\
 2.34741773e-08 & 9.38086308e-01 & 1.45337301e-09 & 2.34741773e-08\\
 2.34741773e-08 & 1.45337301e-09 & 9.38086308e-01 & 2.34741773e-08\\
 1.54929555e-09 & 3.09568451e-02 & 3.09568451e-02 & 9.99999952e-01
\end{bmatrix} \\
\Gamma_{XX} & =
\begin{bmatrix}
9.99999951e-01 & 2.47148265e-02 & 2.47148265e-02 & 1.24580719e-09\\
 2.39578331e-08 & 9.50570344e-01 & 1.18422748e-09 & 2.39578331e-08\\
 2.39578331e-08 & 1.18422748e-09 & 9.50570344e-01 & 2.39578331e-08\\
 1.24580731e-09 & 2.47148287e-02 & 2.47148287e-02 & 9.99999951e-01\\
\end{bmatrix}
\end{aligned}}    
\end{equation}
Note that for the best performance, the $\Gamma_b$ matrices should be experimentally evaluated immediately prior to the VQE experiment.

\begin{figure}
    \centering
    \includegraphics[width = 0.5\textwidth]{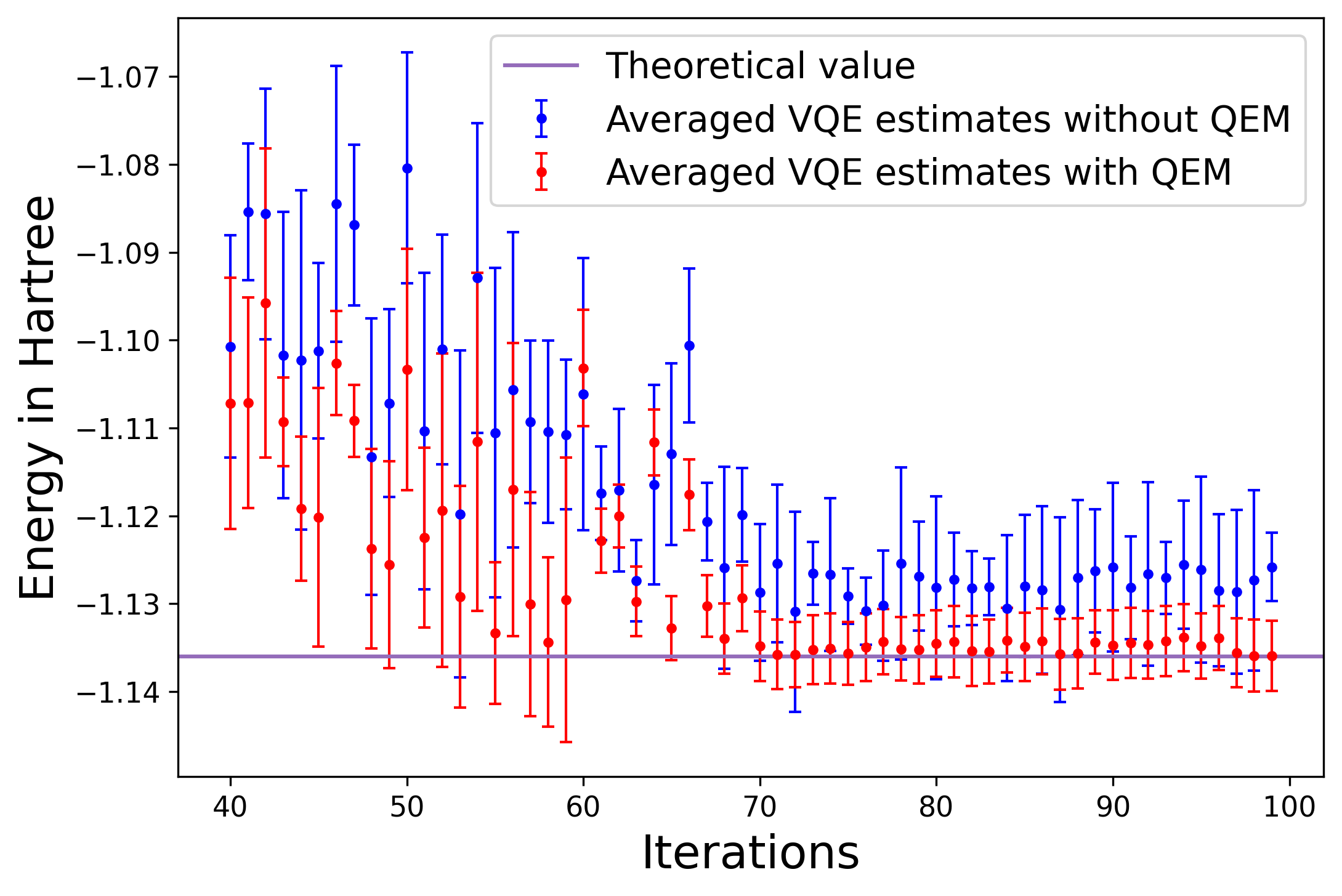}
    \caption{Comparison with and without SPAM error mitigation by averaging loss values over $45$ runs. Error bars are the standard deviation for the average.}
    \label{fig:QEM}
\end{figure}

When we obtain our error-mitigated measurement probabilities, we can construct an eigenvalue estimate which is sent to the classical optimizer. We can note the difference between the expectation values compared to simulated values with and without error mitigation in Fig.~\ref{fig:QEM} and see a noticeable improvement, particularly as the energy nears the ground state.
%Some divergent behaviour is noted in the middle of the optimisation procedure (iterations 20-50) however this can be explained by the optimiser taking different routes towards the global minimum. We see that as the minimum is approached, 
We note that the simulated and error-mitigated values have a close to perfect agreement. This mainly comes from the correction of the basis rotation gates.

\subsection{Classical optimization}

The classical part of the VQE algorithm is performed on the dark blue classical processing unit (CPU) box in Fig.~\ref{fig:procedure_VQE}. The expectation values for the terms of $\hat{\mathcal{H}}_\text{qubit}$ are constructed from error-mitigated samples from the QPU. Then the loss function $\langle \hat{\mathcal{H}}_\text{qubit} \rangle$ is constructed by summing the $5$ terms comprising Eq.~\eqref{eq:qubithamiltonian}. The CPU calls the COBYLA optimizer to perform a gradient descent and find the new set of angles to feed the QPU. Convergence of the procedure is shown in Fig.~\ref{fig:evolution_VQEenergy}.

At each step, we compute the energy from $10000$ processed samples. This amounts to a $0.013$ probability that our sampling error is greater than $0.01$~Hartree.
It is worth noting here that the optimizer can occasionally converge to a local minimum of the objective function, due to the vanishing gradient or barren plateau problem which is avoided in most experiments.

\begin{table}
\small
\begin{tabular}{ |c||c|c|c|c| c |}
 \hline
 \multicolumn{6}{|c|}{Hamiltonian Coefficients} \\
  \hline
 Radius (\AA) & $\alpha$ ($\mathbb{I} \mathbb{I}$)&  $\beta$ ($Z \mathbb{I}$)  &  $\gamma $ ($\mathbb{I} Z$) & $\delta$ ($Z Z$) & $\mu$ ($X X$)\\
 \hline
 \hline
0.2 & 2.0115282039582 & 0.9304885285175 &  0.9304885285175 & 0.013623865138623 & 0.157972708628\\
0.25 & 1.4228278945358 & 0.8706459577114 & 0.8706459577114 & 0.013463487127669  & 0.15927658478468 \\
0.3 & 1.0101820841922 & 0.8086489099089 & 0.8086489099089 & 0.013287977089941 & 0.16081851920392\\
0.35 & 1.0101820841922 & 0.8086489099089 & 0.8086489099089 & 0.013287977089941 & 0.16081851920392 \\
0.4 & 0.4603634956295 & 0.688819429564 & 0.688819429564 & 0.01291396933589 & 0.1645154240225\\
0.45 & 0.2675472248053 & 0.6338897827590 & 0.6338897827590 & 0.012719203005418 & 0.16662140112466 \\
0.50 & 0.11064654485357 & 0.5830796254889 & 0.5830796254889 & 0.01251643158428 & 0.16887022768973\\
0.55 & $-0.0183735206558$ & 0.5364887845888 & 0.5364887845888 & 0.012300353656101 & 0.17124451736495 \\
0.65 & $-0.2139316272136$ & 0.45543342027862 & 0.4554334202786 & 0.011801922101754 & 0.17631845161020\\
0.75 & $-0.3498334175179$ & 0.38874758809160 & 0.38874758809160 & 0.011177144762525 & 0.18177153657730 \\
0.85 & $-0.4454236322275$ & 0.3337464949796 & 0.33374649497965 & 0.01040606826223 & 0.18756184791877 \\
0.95 & $-0.5135484185550$ & 0.2877959899385 & 0.2877959899385 & 0.009503470221825 & 0.19365031698524 \\
1.05 & $-0.5626001130028$ & 0.24878328975518 & 0.24878328975518 & 0.008509936866414 & 0.19998426653596 \\
1.15 & $-0.5979734705198$ & 0.2152339371429 & 0.2152339371429 & 0.007477201225847 & 0.2064946748241 \\
1.25 & $-0.6232232011799$ & 0.1861731031999 & 0.18617310319995 & 0.006455593489009 & 0.2131024013141\\
1.35 & $-0.6408366121165$ & 0.1609263900897 & 0.16092639008976 & 0.005486217221390 & 0.21972703573593 \\
1.45 & $-0.6526612024877$ & 0.13897677941251 & 0.13897677941251 & 0.004597585574594 & 0.22629425934361 \\
1.55 & $-0.6601174612872$ & 0.11989353736336 & 0.11989353736336 & 0.0038055776890 & 0.23274029161766\\
1.65 & $-0.6643091838424$ & 0.10330532972950 & 0.10330532972950 & 0.003115459300252 & 0.23901364608341\\
1.75 & $-0.6660923101667$ & 0.08889055166239 & 0.08889055166239 & 0.00252481503720 & 0.24507502046287\\
1.85 & $-0.6661263822710$ & 0.07637119958665 & 0.07637119958665 & 0.002026489193459 & 0.2508961512677\\
1.95 & $-0.6649159578993$ & 0.06550649596835 & 0.06550649596835 & 0.001610984321150 & 0.2564582470193 \\
2.05 & $-0.6628441004621$ & 0.05608661275595 & 0.05608661275595 & 0.001268117568204 & 0.26175037476834\\
 \hline
\end{tabular}
\caption{List of Hamiltonian coefficients from Eq.~\eqref{eq:qubithamiltonian} for varying bond length.}
\label{tab:Ham_table}
\end{table}
\normalsize

\section{GHZ-Factory}\label{app:ghz}

\begin{figure}[h]
    \centering
    \includegraphics[width=0.7\linewidth]{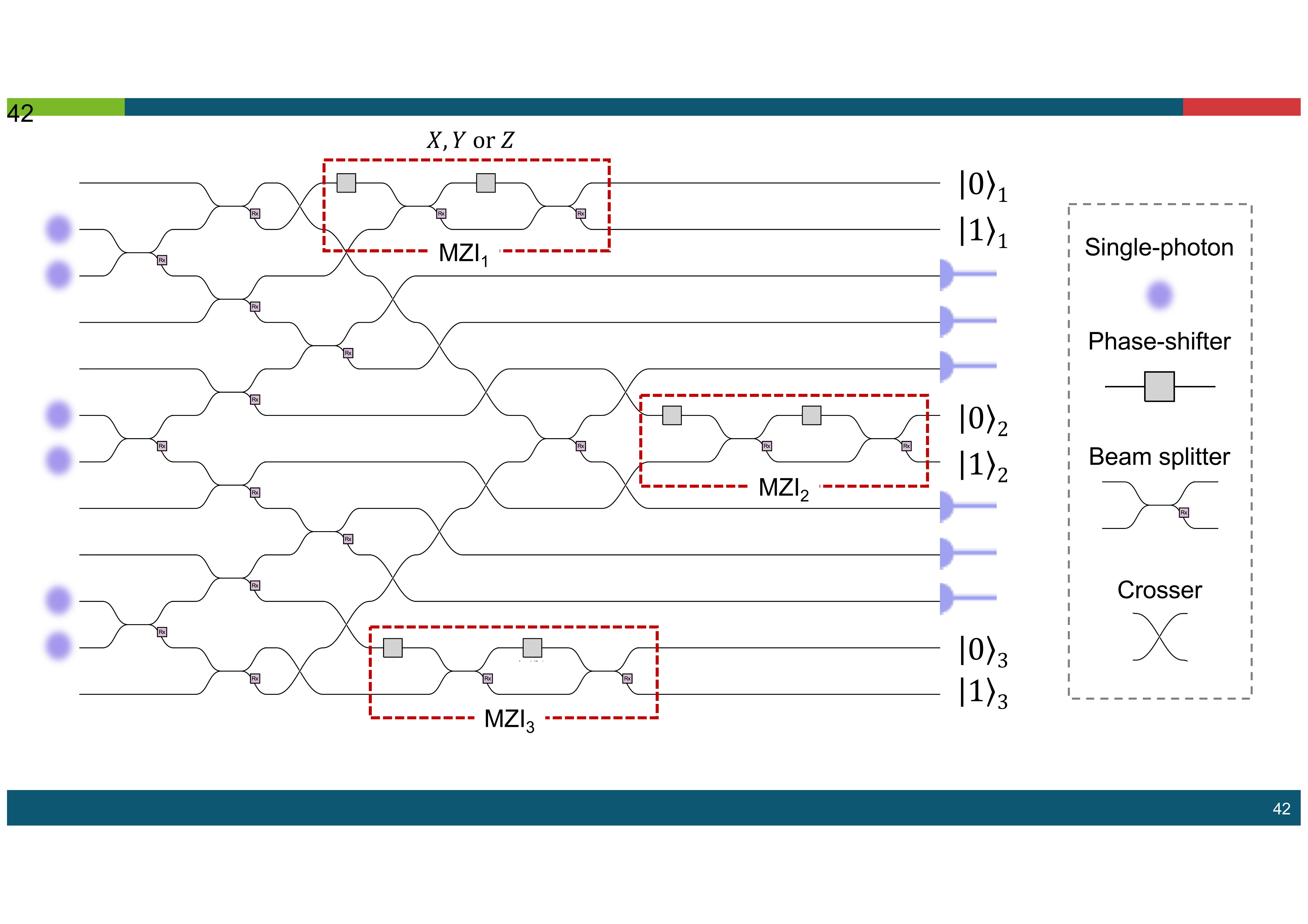}
    \caption{GHZ-factory photonic circuit from \textit{Perceval}. Six single photons (purple dots) are sent to the input optical modes $(2,3,6,7,10,11)$. Three qubits are path encoded in the output optical modes $(1,2)$, $(6,7)$ and $(11,12)$. To account for the losses in the optical system we postselect on the detection of one and only one photon per qubit. We use reconfigurable Mach-Zehnder interferometers to project the generated state in the Pauli matrix basis $(\mathbb{I}, X, Y, Z)$. }
    \label{fig:circuit_3GHZ}
\end{figure}

The layout of the GHZ-factory photonic circuit adapted from Ref.~\cite{li2015,gouriou2019} is presented in Fig.~\ref{fig:circuit_3GHZ}.

\end{document}